\PassOptionsToPackage{prologue,dvipsnames,table}{xcolor}
\documentclass[acmtog,nonacm,table]{acmart}

\usepackage{booktabs} 

\usepackage[ruled]{algorithm2e} 

\SetAlFnt{\small}
\SetAlCapFnt{\small}
\SetAlCapNameFnt{\small}
\SetAlCapHSkip{0pt}

\SetKwComment{Comment}{/* }{ */}

\usepackage[capitalize]{cleveref}
\usepackage{pifont}
\usepackage{overpic}

\usepackage{tikz}
\usepackage{pgfplots}
\usetikzlibrary{positioning,calc,fadings,backgrounds,fit,3d,shapes.misc,shapes.geometric,matrix,hobby}
\pgfplotsset{compat=1.14}

\usepgflibrary{shapes.symbols} 
\usetikzlibrary{shapes.symbols} 

\definecolor{tabcol}{HTML}{a2d2ff}
\usepackage{colortbl}

\usepackage{wrapfig}
\usepackage{fmtcount}

\usepackage{multirow}
\definecolor{textred}{RGB}{209, 71, 111}
\definecolor{textblue}{RGB}{33, 108, 188}
\definecolor{textgreen}{RGB}{0, 126, 0}
\definecolor{textpurple}{RGB}{131, 56, 236}

\definecolor{yellow}{rgb}{1,1, 0.7}
\definecolor{lightyellow}{rgb}{1,1, 0.8}
\definecolor{orange}{rgb}{1, 0.85, 0.7}
\definecolor{tablered}{rgb}{1, 0.7, 0.7}

\newcommand\myqueries{{\textcolor{textgreen}{Queries}}\xspace}
\newcommand\mykeys{{\textcolor{textred}{Keys}}\xspace}
\newcommand\myvalues{{\textcolor{textpurple}{Values}}\xspace}
\newcommand\myweights{{\textcolor{textblue}{Weights}}\xspace}

\newcommand\myquery{{\textcolor{textgreen}{Query}}\xspace}
\newcommand\mykey{{\textcolor{textred}{Key}}\xspace}
\newcommand\myvalue{{\textcolor{textpurple}{Value}}\xspace}
\newcommand\myweight{{\textcolor{textblue}{Weight}}\xspace}

\newcommand{\colorquery}[1]{\textcolor{textgreen}{#1}}
\newcommand{\colorkey}[1]{\textcolor{textred}{#1}}
\newcommand{\colorvalue}[1]{\textcolor{textpurple}{#1}}
\newcommand{\colorweight}[1]{\textcolor{textblue}{#1}}

\newcommand\awesomename{\texttt{e$f$unc}\xspace}

\providecommand{\eg}{\textit{e.g.}\@\xspace}
\providecommand{\ie}{\textit{i.e.}\@\xspace}

\providecommand{\vs}{\textit{v.s.}\@\xspace}

\acmJournal{TOG}

\begin{document}
\title{\texttt{e$f$unc}: An Efficient Function Representation without Neural Networks}

\author[]{Biao Zhang}
\affiliation{
    \institution{KAUST}
    \country{Saudi Arabia}
}
\email{biao.zhang@kaust.edu.sa}

\author[]{Peter Wonka}
\affiliation{
    \institution{KAUST}
    \country{Saudi Arabia}
}
\email{peter.wonka@kaust.edu.sa}

\begin{abstract}
    Function fitting/approximation plays a fundamental role in computer graphics and other engineering applications. While recent advances have explored neural networks to address this task, these methods often rely on architectures with many parameters, limiting their practical applicability. In contrast, we pursue high-quality function approximation using parameter-efficient representations that eliminate the dependency on neural networks entirely. We first propose a novel framework for continuous function modeling. Most existing works can be formulated using this framework. We then introduce a compact function representation, which is based on polynomials interpolated using radial basis functions, bypassing both neural networks and complex/hierarchical data structures. We also develop memory-efficient CUDA-optimized algorithms that reduce computational time and memory consumption to less than 10\% compared to conventional automatic differentiation frameworks.
    Finally, we validate our representation and optimization pipeline through extensive experiments on 3D signed distance functions (SDFs). 
    The proposed representation achieves comparable or superior performance to state-of-the-art techniques (\eg, octree/hash-grid techniques) with significantly fewer parameters.
    
\end{abstract}

%
%


%
%


\begin{teaserfigure}
    \centering
    \includegraphics[width=\textwidth]{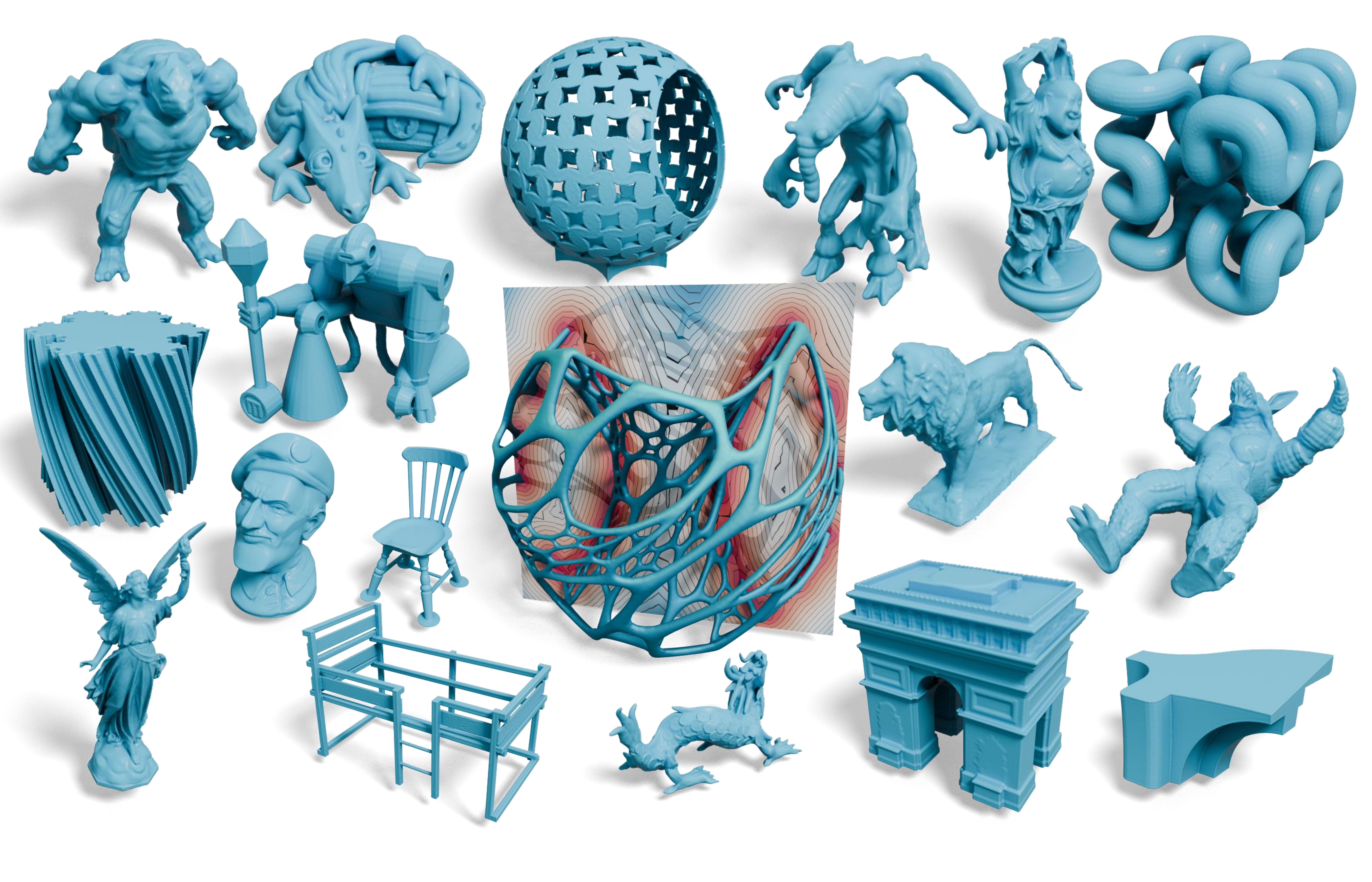}
    \vspace{-30pt}
    \caption{
    We show a collection of shapes represented by \awesomename. Each shape is represented by only $32^3 \times 13$ floats.
    }
    \label{fig:teaser}
\end{teaserfigure}

\maketitle

\section{Introduction}

Function fitting/approximation forms the backbone of modern engineering applications. Within computer graphics, fitting Signed Distance Functions (SDFs) to surfaces is a versatile tool for many downstream applications.

By encoding surfaces implicitly as the signed distance from any point in space to the nearest surface boundary, SDFs support a broad spectrum of applications, including surface reconstruction, shape analysis, collision detection, volumetric rendering, and physical simulations. Their continuous representation of geometry provides significant advantages in terms of compactness, flexibility, and smoothness, making them indispensable for high-quality geometric modeling.
\setlength{\columnsep}{5pt}%
\setlength{\intextsep}{1pt}%
\begin{wrapfigure}{r}{0.5\linewidth}
\centering
\begin{overpic}[trim=5cm 4cm 5cm 5cm,clip,width=1\linewidth,grid=false]{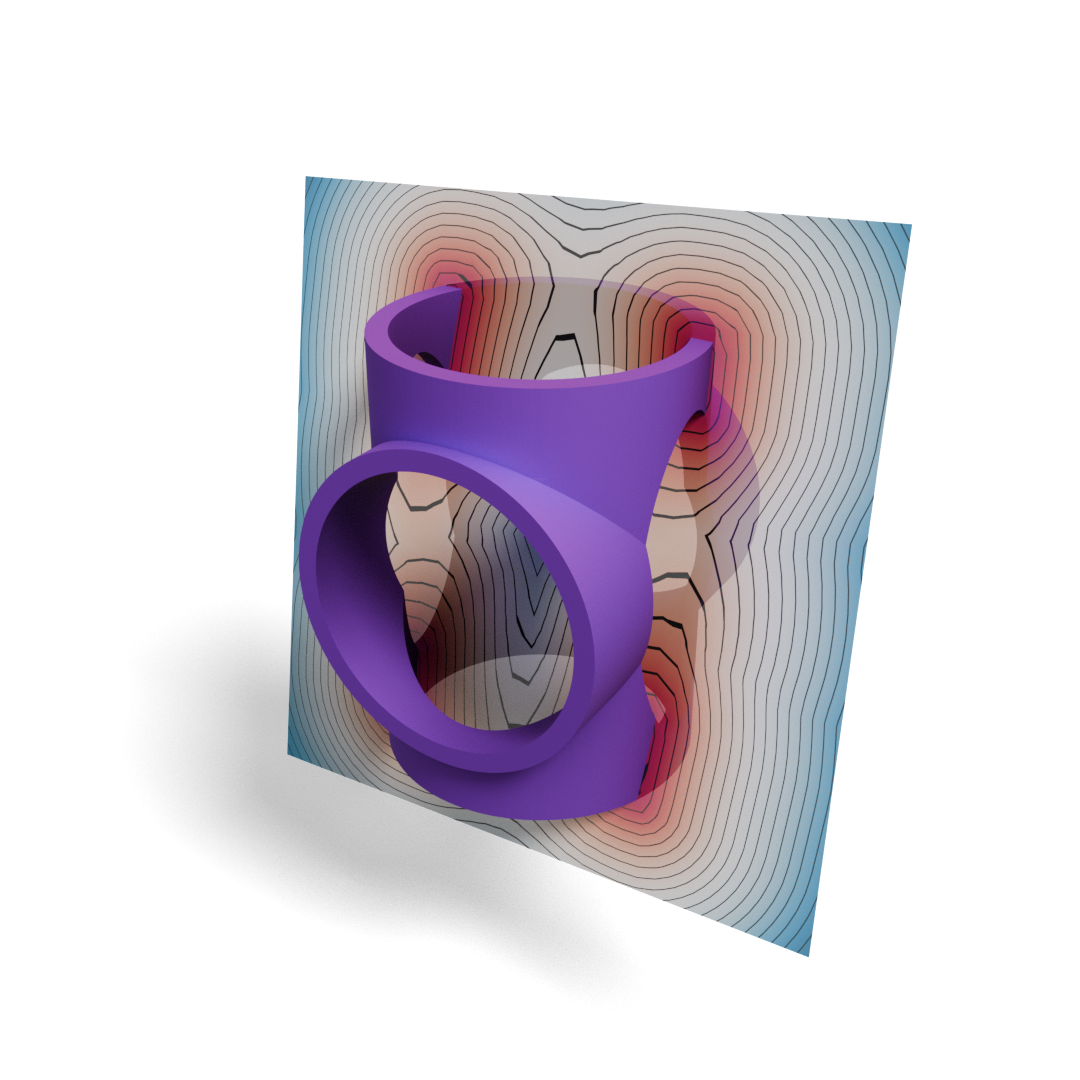}
\end{overpic}\vspace{-3pt}
\end{wrapfigure}

Recent advancements in SDF fitting have been driven by methods leveraging neural networks and specialized data structures, such as octrees~\cite{takikawa2021neural} and hashing grids~\cite{muller2022instant}. These approaches achieve state-of-the-art performance by capitalizing on the high representational capacity of neural models, enabling the accurate fitting of intricate surfaces. However, reliance on sparse or hierarchical data structures introduces additional complexity, complicating integration with other deep learning applications, particularly generative models. 

To overcome these limitations, we propose a novel representation for SDF fitting that eliminates the need for neural networks and complex data structures. We were inspired by recent advancements in differentiable neural rendering~\cite{fridovich2022plenoxels, kerbl20233d}, which demonstrated the potential of non-neural approaches to achieve strong performance. 
Our approach employs regular grids, providing a straightforward and uniform method for data storage and computation. 

A key innovation of this work lies in representing SDFs using simple mathematical functions. Unlike conventional interpolation techniques that focus on scalar or vector values, we utilize function interpolation, where the targets to be interpolated are continuous functions. Our representation is parameterized using radial basis functions and polynomial functions, enabling flexible and efficient encoding of complex surfaces. These parameters are optimized using algorithms commonly employed in deep learning, such as the AdamW optimizer, which effectively minimizes fitting errors.
Notably, our representation achieves high-quality shape reconstruction with a compact storage requirement of only $32^3 \times 13$ floating point values.

The analytic nature of the representation allows for straightforward computation of gradients (for optimization or surface normal estimation), making it particularly advantageous for SDF applications that rely on derivatives, such as normal estimation, Eikonal constraints, and sphere tracing~\cite{hart1996sphere}. This analytic property further enhances the method’s versatility and applicability in geometry processing tasks.

To ensure computational efficiency, we developed a CUDA accelerated framework for both the forward and backward passes of the optimization process. By exploiting the parallel processing capabilities of modern GPUs, this framework significantly accelerates training while enabling fast inference. As a result, the proposed method is well-suited for real-time or near-real-time applications, achieving a balance of simplicity, performance, and efficiency.

In summary, our contributions are as follows:

\begin{itemize}
    \item We introduce a framework for continuous function modeling. Existing work can be analyzed and categorized under this framework.
    \item We propose a novel SDF representation that does not rely on neural networks or complex data structures, utilizing regular grids for simplicity and efficiency.
    \item We develop a CUDA-accelerated computational framework that ensures efficient training and inference.
    \item We demonstrate that our approach achieves comparable performance to state-of-the-art SDF fitting methods while being significantly with an order of magnitude fewer parameters.
\end{itemize}

\begin{figure}
    \centering
    \input{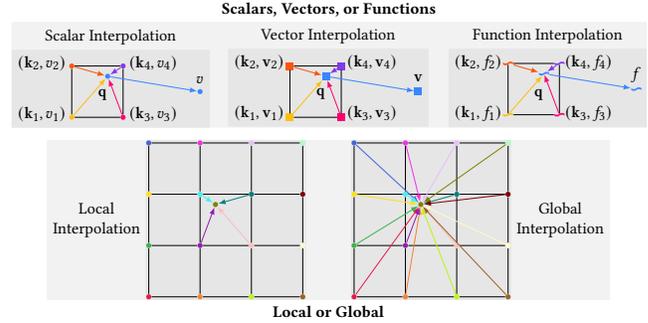}
    \vspace{-20pt}
    \caption{We analyze how we interpolate. The target can be scalars, vectors, or functions (\emph{top}). The support can be local or global (\emph{bottom}). In this work, we are using functions as the interpolating targets with global support. }
    \label{fig:interp}
\end{figure}

\section{Continuous Function Modeling}
In this section, we propose a novel framework to analyze and categorize previous work. We build our framework by first introducing a basic formulation in~\cref{sec:func-approx} and then extending it to the full framework in~\cref{sec:practical}. We will also describe how most previous work can be mapped into this framework.  

\subsection{Basic Framework}\label{sec:func-approx}
Continuous functions can be approximated using simpler functions (\textit{e.g.}, polynomials, exponentials, and sinusoids) as building blocks. We propose to represent
a function $\mathcal{F}:\mathbb{R}^{\mathrm{in}}\rightarrow \mathbb{R}^{\mathrm{out}}$ in the following form 
\begin{equation}\label{eq:func-approx}
    \mathcal{F}(\textcolor{textgreen}{\mathbf{q}}) = \sum_{i} \textcolor{textblue}{w}(\textcolor{textgreen}{\mathbf{q}}, \textcolor{textred}{\mathbf{k}_i})
     \cdot \textcolor{textpurple}{\mathbf{v}_i},
\end{equation}
with the variables 
\begin{itemize}
    \item $\mathbf{q}$: \textcolor{textgreen}{Queries}, which lie in the space $\mathbb{R}^{\mathrm{in}}$;
    \item $\mathbf{k}_i$: \textcolor{textred}{Keys}, which lie in the same space $\mathbb{R}^{\mathrm{in}}$;
    \item $\mathbf{v}_i$: \textcolor{textpurple}{Values}, which lie in the space of $\mathbb{R}^{\mathrm{out}}$.
    \item $w(\cdot, \cdot)$: \textcolor{textblue}{Weights}, which are scalar-valued weight functions $\mathbb{R}^{\mathrm{in}}\times \mathbb{R}^{\mathrm{in}}\rightarrow\mathbb{R}$.
\end{itemize}

We use capitalized terms \myqueries, \mykeys, \myvalues, and \myweights to avoid ambiguity.
Real-valued function approximations such as Taylor series, kernel approximations, and Fourier series can also be expressed in this form (see~\cref{tab:func-approx} for details),
\begin{equation}
    \begin{aligned}
        \text{Taylor series}: && \mathcal{F}(\textcolor{textgreen}{q})&=\sum_i \textcolor{textblue}{\mathrm{pow}}(\textcolor{textgreen}{q}, \textcolor{textred}{i})\cdot \textcolor{textpurple}{v_i};\\
        \text{Fourier cosine series}: && \mathcal{F}(\textcolor{textgreen}{q})&=\sum_i \textcolor{textblue}{\cos}\left(\textcolor{textred}{2\pi i/P}\cdot\textcolor{textgreen}{q}\right)\cdot \textcolor{textpurple}{v_i}; \\
        \text{Kernel approximations}: && \mathcal{F}(\textcolor{textgreen}{q})&=\sum_i \textcolor{textblue}{\phi}\left(\left\|\textcolor{textgreen}{\mathbf{q}}-\textcolor{textred}{\mathbf{k}_i}\right\|\right)\cdot \textcolor{textpurple}{\mathbf{v}_i}. \\
    \end{aligned}
\end{equation}
Note that the \myweights can be negative. In mathematical analysis, they are often termed basis functions.

\subsubsection{Support}
\paragraph{Local support.} Local interpolation methods query a point's function value by considering its neighboring points $\sum_{i\in\mathcal{N}(\colorquery{\mathbf{q}})} \textcolor{textblue}{w}(\textcolor{textgreen}{\mathbf{q}}, \textcolor{textred}{\mathbf{k}_i})
     \cdot \textcolor{textpurple}{\mathbf{v}_i}$, where $\mathcal{N}(\colorquery{\mathbf{q}})$ represents the neighborhood of $\colorquery{\mathbf{q}}$. Many methods fall into this category, including~\cite{yan2022shapeformer, peng2020convolutional, ren2024xcube, yariv2024mosaic, takikawa2021neural, muller2022instant}. However, because 3D data is often spatially sparse, these methods require high-resolution grids or specially designed data structures to capture fine details (\eg, octrees). Another problem is that, building the neighborhood set $\mathcal{N}(\colorquery{\mathbf{q}})$ often necessitates fast algorithms for neighborhood search (\eg, k-d tree). Additionally, we need to tune a hyperparameter, \ie, the size of the neighborhood.

\paragraph{Global support.} In contrast, global support leverages all known data points to determine the function value at a query point,
    $\sum_{i}^I \textcolor{textblue}{w}(\textcolor{textgreen}{\mathbf{q}}, \textcolor{textred}{\mathbf{k}_i})
     \cdot \textcolor{textpurple}{\mathbf{v}_i}$.
This approach maximizes information usage by ensuring that each known element contributes to the interpolation. 3DILG~\cite{zhang20223dilg} uses irregular grids and interpolates over all features, while VecSet~\cite{zhang20233dshape2vecset} interprets Attention~\cite{vaswani2017attention} as a learnable form of global interpolation. While global interpolation fully utilizes available information, it also has drawbacks, including high time and memory complexity. See~\cref{fig:interp} (bottom) for an illustration.

\subsubsection{Interpolating targets}
The scalar- or vector-\myvalues in~\cref{eq:func-approx} often have limited representational capacity.
To enhance the expressive power,
we can further adapt~\cref{eq:func-approx} to
\begin{equation}\label{eq:func-approx-func}
    \mathcal{F}(\textcolor{textgreen}{\mathbf{q}}) = \sum_{i} \textcolor{textblue}{w}(\textcolor{textgreen}{\mathbf{q}}, \textcolor{textred}{\mathbf{k}_i})
     \cdot \textcolor{textpurple}{\mathbf{f}}(\textcolor{textgreen}{\mathbf{q}}, \textcolor{textred}{\mathbf{k}_i}; \boldsymbol{\phi}_i),
\end{equation}
where we replace the \myvalue with a function $f:\mathbb{R}^{\text{in}}\times \mathbb{R}^{\text{in}} \times \mathbb{R}^{|\boldsymbol{\phi}|} \rightarrow \mathbb{R}^{\text{out}}$. The \myvalue $f$ is parameterized by $\boldsymbol{\phi}_i$. Also see~\cref{fig:interp} (top) for visualizations.
With this formulation, we hope that the \myvalue $\colorvalue{f}$ contains more information than either a simple scalar $\colorvalue{v}$ or a vector $\colorvalue{\mathbf{v}}$.

\paragraph{Scalar Interpolation.} Simple bi/trilinear interpolation and early methods~\cite{carr2001reconstruction} employ scalar interpolation for SDF modeling. While scalar interpolation is simple and does not require neural network parameters, it often exhibits limited expressive power.

\paragraph{Vector Interpolation.} Most neural network-based methods fall into this category~\cite{muller2022instant, takikawa2021neural, peng2020convolutional, yan2022shapeformer, zhang20223dilg, zhang20233dshape2vecset}. These approaches typically query a high-dimensional feature vector through interpolation, which is then decoded using MLPs. Combining interpolation with MLPs generally provides better expressiveness compared to simple scalar interpolation.

\paragraph{Function Interpolation.} To address the limitations of scalar interpolation while retaining the expressiveness of MLP decoders, some works utilized function interpolation. It has been explored in early works such as MPU~\cite{ohtake2023mpu} and IMLS~\cite{kolluri2008provably}.
M-SDF~\cite{yariv2024mosaic} can also be interpreted as function interpolation by treating trilinear grid interpolation as the \myvalue.

\begin{table}[tb]
    \centering
    \caption{Framework for function approximations.}
    \label{tab:type-func-approx}
    \begin{tabular}{ccc}
    \toprule
        $\mathcal{O}=\mathcal{D}\circ\mathcal{F}$ & $\mathcal{D}$ & $\mathcal{F}$\\ \midrule
        Type-I & \textcolor{gray!70}{Identity function $\mathcal{I}$} & \cref{eq:func-approx} or \cref{eq:func-approx-func} \\
        Type-II & Network $\mathrm{MLP}(\cdot)$ & \textcolor{gray!70}{Identity function $\mathcal{I}$} \\
        Type-III & Network $\mathrm{MLP}(\cdot)$  & \cref{eq:func-approx} \\
    \bottomrule
    \end{tabular}
\end{table}

\begin{table*}[htbp]
    \centering
    \def\arraystretch{1}\tabcolsep=0.35em 
    
    \caption{\textbf{Function representations.} We show a categorization of related work. There are some other works that share similar methodologies that are omitted here for clarity. For example, DeepSDF~\cite{park2019deepsdf} is similar to ONet~\cite{mescheder2019occupancy}; 
    ReLU Fields~\cite{karnewar2022relu}, Plenoxels~\cite{fridovich2022plenoxels}, and DVGO~\cite{sun2022direct} follow Trilinear Interpolation; PointNerf~\cite{xu2022point} follows a similar design as~\citet{carr2001reconstruction};  GALA~\cite{yang2024gala} follows a similar formulation as M-SDF~\cite{yariv2024mosaic}.}
    \label{tab:func-approx}
        \begin{tabular}{cc>{\color{textred}}c>{\color{textpurple}}c>{\color{textpurple}}c>{\color{textblue}}c>{\color{textblue}}cccc}
    \toprule
    & \multirow{3}{*}{Methods} & \multicolumn{5}{c}{$\mathcal{F}$} & $\mathcal{D}$ & \multirow{3}{*}{
        \begin{tabular}{c}
            Per-\\
            Object  \\
            Fitting
        \end{tabular}
    } & \multirow{3}{*}{
        \begin{tabular}{c}
            Per-\\
            Dataset  \\
            Fitting
        \end{tabular}
    }\\
    \cmidrule(lr){3-7}\cmidrule(lr){8-8}
     & & \multirow{2}{*}{Keys} & \multicolumn{2}{c}{\textcolor{textpurple}{Values}} & \multicolumn{2}{c}{\textcolor{textblue}{Weights}} & \multirow{2}{*}{Decoding} &  \\
     \cmidrule(lr){4-5}\cmidrule(lr){6-7}
    & & & Type & Function & Function &  Support &  & \\ 
        \midrule
        \midrule
        & Taylor series & Integers & Scalars & - & Power & Global & - & - & -\\
        & Kernel approximation & Spatial Points & Scalars & - & RBF & Global & - & - & -\\
        & Fourier series & Frequencies & Scalars & - & Trigonometric & Global & - & -  & -\\
        \midrule
        \multirow{7}{*}{\rotatebox[origin=c]{90}{Type-I}} & Bi/Trilinear Interpolation & Grid & Scalars & - & Trilinear & Local & Identity & \ding{51} & \ding{51} \\
        & \citet{carr2001reconstruction} & Surface$^*$ & Scalars & - & RBF & Local & Identity & \ding{51} & \ding{55} \\ 
        & MPU~\cite{ohtake2023mpu} & Octree & Functions & Quadratic & B-Spline & Local & Identity & \ding{51} & \ding{55} \\
        & IMLS~\cite{kolluri2008provably} & Surface & Functions & Linear & RBF & Local & Identity & \ding{51} & \ding{55} \\ 
        & M-SDF~\cite{yariv2024mosaic} & Surface & Functions & Trilinear & Triangular & Local & Identity & \ding{51} & \ding{55} \\
        & DMP~\cite{he2024dmp} & Surface$^*$ & Functions & Quadratic & B-Spline & Local & Identity & \ding{51} & \ding{55} \\
        & CoFie~\cite{jiang2024cofie} & Sparse Grid & Functions & Quadratic & Constant & Global & Identity & \ding{55} & \ding{51}\\
        \midrule
        \multirow{3}{*}{\rotatebox[origin=c]{90}{Type-II}}& ONet~\cite{mescheder2019occupancy} & \multicolumn{5}{c}{\multirow{3}{*}{Identity}} & \multirow{3}{*}{MLP} & \ding{55} & \ding{51} \\
        & SIREN~\cite{sitzmann2020implicit} & & & & & & & \ding{51} & \ding{55}\\
        & FFN~\cite{tancik2020fourier} & & & & & & & \ding{51} & \ding{55}\\
        \midrule
        \multirow{6}{*}{\rotatebox[origin=c]{90}{Type-III}} & C-ONet~\cite{peng2020convolutional} & Grid & Vectors & - & Trilinear & Local & MLP & \ding{55} & \ding{51} \\
        & NGLOD~\cite{takikawa2021neural} & Octree & Vectors & - & Trilinear & Local & MLP & \ding{51} & \ding{55} \\
        & I-NGP~\cite{muller2022instant} & Hashing Grid & Vectors & - & Trilinear & Local & MLP & \ding{51} & \ding{55} \\ 
        & 3DILG~\cite{zhang20223dilg} & Surface & Vectors & - & RBF & Global & MLP & \ding{55} & \ding{51} \\
        & VecSet~\cite{zhang20233dshape2vecset} & Learnable & Vectors & - & Learnable & Global & MLP & \ding{55} & \ding{51} \\
        & DiF-Grid~\cite{chen2023dictionary} & Grid & Vectors & - & Trilinear & Local & MLP & \ding{51} & \ding{55}\\
        \midrule
        \midrule
        & Ours & Grid & Functions & Polynomial & RBF & Global & Identity & \ding{51} & \ding{55} \\
        \bottomrule
        \multicolumn{10}{l}{$^*$ Off-surface points might also be used in the representations.}
    \end{tabular}
\end{table*}

\subsection{Main Framework}\label{sec:practical}
While our framework is more general, we explain it using signed distance functions (SDFs) as the main application.
For a given surface, an SDF, denoted as  $\mathcal{O}(\mathbf{q}):\mathbb{R}^3\rightarrow\mathbb{R}$, represents the distance from a \myquery $\mathbf{q}\in\mathbb{R}^3$ (a spatial point) to the closest point on the surface. We propose the following as our main framework,
\begin{equation}
\mathcal{O}(\textcolor{textgreen}{\mathbf{q}})=\mathcal{D}(\mathcal{F}(\textcolor{textgreen}{\mathbf{q}})),
\end{equation}
where $\mathcal{F}$ is for function representation as in~\cref{sec:func-approx} and $\mathcal{D}$ is for projection (or decoding). We structure related work into three different types. An overview of the types can be found in~\cref{tab:type-func-approx}. The addition of the decoder $\mathcal{D}$ helps to integrate MLPs into the framework from the previous section.


\paragraph{Type-I} A naive approach to represent SDFs involves using a grid of \myvalues and applying trilinear interpolation given an arbitrary \textcolor{textgreen}{Query} point,
\begin{equation}\label{eq:trilinear}
\mathcal{O}^{\text{trilinear}}(\textcolor{textgreen}{\mathbf{q}})=\sum_{i\in\mathcal{N}(\textcolor{textgreen}{\mathbf{q}})}\textcolor{textblue}{\mathrm{Trilinear}}(\textcolor{textgreen}{\mathbf{q}}, \textcolor{textred}{\mathbf{k}_i})\cdot \textcolor{textpurple}{v_i}.
\end{equation}
Early works modeled the SDFs with a finite sum that can be unified into the framework as in~\cref{eq:func-approx}. We denote it as 
\begin{equation}
\mathcal{O}(\textcolor{textgreen}{\mathbf{q}})=\mathcal{I}(\mathcal{F}(\textcolor{textgreen}{\mathbf{q}}))=\mathcal{F}(\textcolor{textgreen}{\mathbf{q}}),
\end{equation}
where $\mathcal{D}=\mathcal{I}$ is an identity function.
For example, \citet{carr2001reconstruction}'s formulation matches kernel approximations except that the \myweights have local support. Other methods, such as MPU~\cite{ohtake2023mpu} and IMLS~\cite{kolluri2008provably}, employ functions as the \myvalues as in~\cref{eq:func-approx-func}. 
To achieve strong representational power, these classical methods often require the \mykeys to number in the millions~\cite{carr2001reconstruction, he2024dmp} or reply on complex data structures (such as octrees in~\cite{ohtake2023mpu, wang2022dual}). 
While M-SDF~\cite{yariv2024mosaic} used a far smaller number of \mykeys, its \myvalues are parameterized by high-dimensional features.

\paragraph{Type-II} Over the recent decade, there has been a growing trend~\cite{park2019deepsdf, tancik2020fourier, sitzmann2020implicit} to represent functions using neural networks -- specifically MLPs of the form $\mathrm{MLP}:\mathbb{R}^3\rightarrow\mathbb{R}$. 
The methods are typically expressed as
\begin{equation}
\mathcal{O}(\textcolor{textgreen}{\mathbf{q}})=\mathrm{MLP}(\mathcal{I}(\textcolor{textgreen}{\mathbf{q}}))=\mathrm{MLP}(\textcolor{textgreen}{\mathbf{q}}),
\end{equation}
where $\mathcal{D}=\mathrm{MLP}$ and $\mathcal{F}=\mathcal{I}$.
However, they often suffer from a limited representation capability and slow convergence.

\paragraph{Type-III} Recent methods~\cite{peng2020convolutional, takikawa2021neural, muller2022instant, zhang20223dilg, zhang20233dshape2vecset} enhance neural networks by incorporating local data structures such as regular grids~\cite{peng2020convolutional}, octrees~\cite{takikawa2021neural}, hashing grids~\cite{muller2022instant}, irregular grids~\cite{zhang20223dilg}, and vector sets~\cite{zhang20233dshape2vecset}. These approaches can also be unified into the function representation framework in~\cref{eq:func-approx}. Unlike classical methods~\citep{carr2001reconstruction, ohtake2023mpu, kolluri2008provably}, these neural-network-based methods interpolate high-dimensional vectors $\mathbb{R}^{\mathrm{out}}$ (features or latents) and use a lightweight MLP to decode them into scalars, 
\begin{equation}
\mathcal{O}(\textcolor{textgreen}{\mathbf{q}})=\mathrm{MLP}(\mathcal{F}(\textcolor{textgreen}{\mathbf{q}})),
\end{equation}
where $\mathcal{D}=\mathrm{MLP}$.
Overall, these methods mixing neural networks and local data structures often achieve state-of-the-art performance with a reasonable number of parameters.

A summary of the detailed designs can be found in~\cref{tab:func-approx}.

\section{Method}

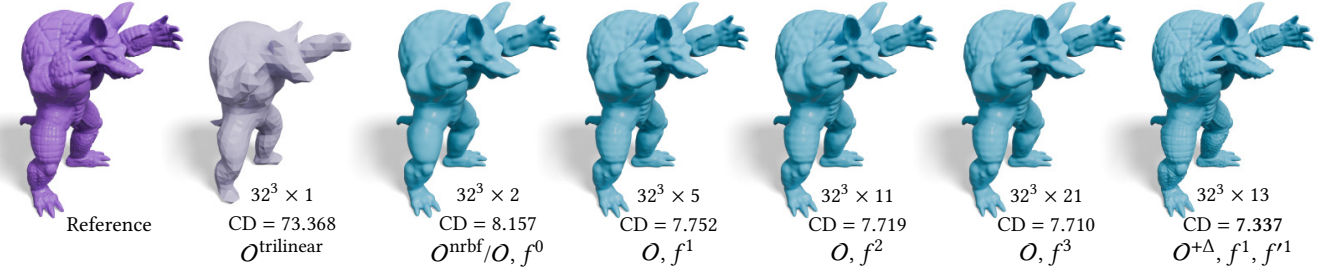
\begin{figure*}[htb]
    \begin{overpic}[trim=2cm 0cm 1.5cm 0cm,clip,width=1\linewidth,grid=false]{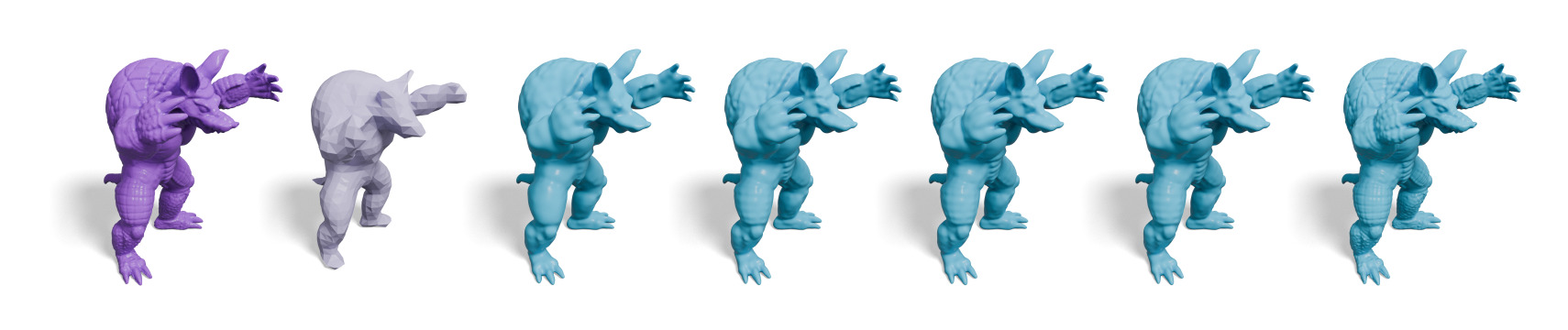}

    \put (18,2) {\begin{tabular}{c}  
        \small $32^3 \times 1$\\ \small $\mathrm{CD}=73.368$ \\ $\mathcal{O}^{\text{trilinear}}$ 
    \end{tabular}}

    \put (33,2) {\begin{tabular}{c}  
        \small $32^3 \times 2$\\ \small $\mathrm{CD}=8.157$ \\ $\mathcal{O}^{\text{nrbf}}$/$\mathcal{O}$, $f^0$
    \end{tabular}}
    
    \put (47,2) {\begin{tabular}{c}  
        \small $32^3 \times 5$\\ \small $\mathrm{CD}=7.752$  \\ $\mathcal{O}$, $f^1$
    \end{tabular}}

    \put (61,2) {\begin{tabular}{c}  
        \small $32^3 \times 11$\\ \small $\mathrm{CD}=7.719$  \\ $\mathcal{O}$, $f^2$
    \end{tabular}}

     \put (75,2) {\begin{tabular}{c}  
        \small $32^3 \times 21$\\ \small $\mathrm{CD}=7.710$  \\ $\mathcal{O}$, $f^3$
    \end{tabular}}

    \put (88,2) {\begin{tabular}{c}  
        \small $32^3 \times 13$\\ \small $\mathbf{\mathrm{CD}=7.337}$  \\ $\mathcal{O}^{+\Delta}$, $f^1$, $f'^1$
    \end{tabular}}
    
    \put (7,2) {\small Reference}

    \end{overpic}
    \vspace{-15pt}
    \caption{Quality of the representations. We start from a simple trilinear method applied on a 32-resolution grid (\emph{gray}). We show the results obtained using the proposed representations (\emph{blue}). The ground-truth mesh is shown on the left (\emph{purple}) as a reference. The best results are obtained using the proposed $\mathcal{O}^{+\Delta}$ in~\cref{eq:func-offset}. For the other figures, we can see that the results are getting better when using a higher order of $f$. However, the differences in visual results are not as significant as the metrics.}
    \label{fig:intro}
\end{figure*}

Building on this function representation framework, we propose an SDF representation that achieves higher accuracy for a given parameter budget while retaining a grid structure.
First, we begin with trilinear and RBF interpolations in~\cref{sec:tri-rbf}.
Second, we introduce polynomial-based \myvalue functions in~\cref{sec:poly}.
Third, we enhance representational capacity near 3D object surfaces through offsets in~\cref{sec:mean-shift}.
Fourth, we demonstrate that the loss function can be optimized using deep learning-inspired techniques in~\cref{sec:fitting}.
Finally, we present forward and backward algorithms for the representation in~\cref{sec:forward} and~\cref{sec:backward}, respectively. An overview of the fitting results can be found in~\cref{fig:intro}.

\subsection{Basic formulation}\label{sec:tri-rbf}

We start from the simplest form as in~\cref{eq:trilinear}.
To represent high-quality, detailed objects, a large grid resolution $R$ (for example, 512, 1024, or even higher) is required. However, the cubic storage complexity makes the representation impractical for many applications. 
Existing works to address the issue include octrees~\cite{ohtake2023mpu,wang2022dual, takikawa2021neural}, sparse point clouds~\cite{carr2001reconstruction,kolluri2008provably,zhang20223dilg}, 
and hashing grids~\cite{muller2022instant}. 
Our goal is to design data structures that are well-defined and seamlessly integrate with neural networks. Despite the challenges, a regular grid-like structure remains the most promising choice.

The limitation of~\cref{eq:trilinear}
lies in its dependence on only the 8 nearest grid points (or 4 neighbors in 2D) to compute the signed distance.
The localized dependency is inherently inefficient.
The \mykeys are only active within a small region in this case.
A straightforward modification to address this issue is to allow each \mykey (grid point) to influence a larger region. We consider the following representation,
\begin{equation}
    \mathcal{O}^{\text{rbf}}(\textcolor{textgreen}{\mathbf{q}}; \left\{\textcolor{textred}{\mathbf{k}_i}, \beta_i, \textcolor{textpurple}{v_i}\right\}_{i=1}^I) 
    = \sum^I_i\left[\textcolor{textblue}{\exp}(-\beta_i\left\|
        \textcolor{textgreen}{\mathbf{q}}-\textcolor{textred}{\mathbf{k}_i}
    \right\|^2) \cdot \textcolor{textpurple}{v_i}\right].
\end{equation}
The parameters $\boldsymbol{\theta} = \{\mathbf{k}_i, \beta_i, v_i\}_{i=1}^{I}\in\mathbb{R}^{R\times R\times R \times 2}$ and $I=R^3$.
In this formulation,
\begin{enumerate}
    \item $1/\beta_i$ determines the radius of the affecting region of a \mykey $\colorkey{\mathbf{k}_i}$ (grid point).
    \item $\colorvalue{v_i}$ is the \myvalue assigned to $\colorkey{\mathbf{k}_i}$.
    \item The influence of a \mykey decreases with the distance between the \myquery $\colorquery{\mathbf{q}}$ and $\colorkey{\mathbf{k}_i}$, as governed by the \myweight function $\colorweight{\exp}\left(-\beta_i\left\|\colorquery{\mathbf{q}}-\colorkey{\mathbf{k}_i}\right\|^2\right)$
\end{enumerate}
To further improve stability during training, we normalize the \myweights using the following formulation:
\begin{equation}\label{eq:nrbf}
\begin{aligned}
    \mathcal{O}^{\text{nrbf}}(\colorquery{\mathbf{q}}; \left\{\colorkey{\mathbf{k}_i}, \beta_i, \colorvalue{v_i}\right\}_{i=1}^I) &= 
    \sum^I_i\left[\frac{1}{Z}\colorweight{\exp}(-\beta_i\left\|\colorquery{\mathbf{q}}-\colorkey{\mathbf{k}_i}\right\|^2) \cdot \colorvalue{v_i}\right],\\
    &= 
    \sum^I_i\left[\colorweight{\mathrm{softmax}}(-\beta_i\left\|\colorquery{\mathbf{q}}-\colorkey{\mathbf{k}_i}\right\|^2) \cdot \colorvalue{v_i}\right]
\end{aligned}
\end{equation}
where $Z=\sum^I_i\exp(-\beta_i\left\|\colorquery{\mathbf{q}}-\colorkey{\mathbf{k}_i}\right\|^2)$ is a normalizing factor.
This normalization ensures numerical stability during the training process and has been widely adopted in machine learning (\textit{e.g.}, Nadaraya-Watson kernel regression), where it is often referred to as the \emph{softmax} function. 
The function is a convex combination (interpolation) of the \myvalues $\{\colorvalue{v_i}\}_{i}^I$, effectively blending contributions from all \mykeys $\{\colorkey{\mathbf{k}_i}\}_{i}^I$. This formulation only introduces one additional parameter per grid cell, compared to trilinear interpolation.
We now build a radial basis function representation where each radial basis function is centered around a grid point.

\subsection{Extension: Functions instead of Scalars}\label{sec:poly}
We extend the representation further by replacing the scalar \myvalue $\colorvalue{v_i}$ at each \mykey $\colorkey{\mathbf{k}_i}$ with a scalar function $f(\colorquery{\mathbf{q}}, \colorkey{\mathbf{k}_i};\boldsymbol{\phi}_i)\in\mathbb{R}$, parameterized by $\boldsymbol{\phi}_i$.
This approach enhances the expressive power of the representation compared to using a simple scalar value (also see~\cref{eq:func-approx-func}). Conceptually, this modification enables us to interpolate \emph{functions} rather than scalars, resulting in the following formulation:
\begin{equation}\label{eq:func-interp}
    \mathcal{O}(\colorquery{\mathbf{q}}; \left\{\colorkey{\mathbf{k}_i}, \beta_i, \boldsymbol{\phi}_i\right\}_{i=1}^I) = \sum^I_i\left[\colorweight{\mathrm{softmax}}(-\beta_i\left\|\colorquery{\mathbf{q}}-\colorkey{\mathbf{k}}_i\right\|^2) \colorvalue{f}(\colorquery{\mathbf{q}}, \colorkey{\mathbf{k}_i};\boldsymbol{\phi}_i)\right].
\end{equation}
The parameter space of the representation is now expanded to $\mathbb{R}^{R^3 \times (1+|\boldsymbol{\phi}|)}$.
The design of $\colorvalue{f}$ offers significant flexibility, with the goal of balancing simplicity and expressive power. Additionally, we impose the property of \emph{shift-invariance} of $\colorvalue{f}$, requiring that it depend only on the relative vector $\colorquery{\mathbf{q}}-\colorkey{\mathbf{k}_i}$. This ensures that $f$ can be expressed as $\colorvalue{f}(\colorquery{\mathbf{q}}-\colorkey{\mathbf{k}_i};\boldsymbol{\phi}_i)$.

To model $\colorvalue{f}$, one could employ a multilayer perceptron (MLP). However, achieving high expressiveness with an MLP would introduce an excessive number of parameters, complicating the representation~\cite{takikawa2021neural,muller2022instant,chen2023dictionary}. A simpler alternative is to use a polynomial function:
\begin{equation}\label{eq:poly-func}
    f(\mathbf{x};\boldsymbol{\phi}) = \circ + \mathbf{x}^\intercal \diamond + \frac{1}{2}\mathbf{x}^\intercal\square\mathbf{x}+\cdots,
\end{equation}
where $\boldsymbol{\phi}=\{\circ\in\mathbb{R}, \diamond\in\mathbb{R}^3, \square\in\mathbb{R}^{3\times 3}, \cdots\}$ parameterizes the function. This formulation aligns with the Taylor series, allowing the function to approximate any sufficiently smooth function as the order increases, according to Taylor's theorem in calculus.
In practice, a truncated polynomial is used to approximate $\colorvalue{f}$. Examples include:
\begin{itemize}
    \item 0th order: $f^0(\mathbf{x};\boldsymbol{\phi}) = \circ$,
    \item 1st order: $f^1(\mathbf{x};\boldsymbol{\phi}) = \circ + \mathbf{x}^\intercal \diamond$,
    \item 2nd order: $f^2(\mathbf{x};\boldsymbol{\phi}) = \circ + \mathbf{x}^\intercal \diamond + \frac{1}{2}\mathbf{x}^\intercal\square\mathbf{x}$.
\end{itemize}
When using the 0th-order approximation, the representation reduces to the normalized radial basis function in~\cref{eq:nrbf}. The 1st-order approximation introduces a linear dependency on $\colorquery{\mathbf{q}}-\colorkey{\mathbf{k}_i}$, providing a more expressive representation while maintaining simplicity. 
Classical works such as MPU~\cite{ohtake2023mpu} (quadratic functions) and IMLS~\cite{kolluri2008provably} (linear functions) also interpolate polynomial functions, but they require a lot more parameters than our representation for the same approximation quality. The main reason is the small local support of the radial basis functions they choose to interpolate the polynomials, and the lack of offsets for MPUs. Furthermore, \citet{yariv2024mosaic} employed a special type of polynomial, trilinear grids. In this case, the \myvalue can be expressed as
$$f^{\text{cube}}(\mathbf{x})=a_0+a_1x_1+a_2x_2+a_3x_3+a_4x_1x_2+a_5x_1x_3+a_6x_2x_3+a_7x_1x_2x_3,
$$
where $a_{0:7}$ are trilinear coefficients.
The formulation is also a special case of~\cref{eq:poly-func} (missing some higher-order terms).


\paragraph{Gradients of \myqueries.} The gradient of $\mathcal{O}(\colorquery{\mathbf{q}})$ with respect to $\colorquery{\mathbf{q}}$ can be derived as,
\begin{equation}\label{eq:func-normal}
    \frac{\partial \mathcal{O}(\colorquery{\mathbf{q}})}{\partial \colorquery{\mathbf{q}}} = 
        \sum^I_i \left[
            \colorweight{\mathrm{softmax}}\left(-\beta_i\left\|\colorquery{\mathbf{q}} -\colorkey{\mathbf{k}_i}\right\|^2\right) g(\colorquery{\mathbf{q}}-\colorkey{\mathbf{k}_i})
        \right]
\end{equation}
where 
\begin{equation}
    \begin{aligned}
        g(\colorquery{\mathbf{q}}-\colorkey{\mathbf{k}_i}) = \frac{\partial f(\colorquery{\mathbf{q}}-\colorkey{\mathbf{k}_i};\boldsymbol{\phi}_i)}{\partial \colorquery{\mathbf{q}}} +2\cdot\beta_i\cdot(\colorquery{\mathbf{q}}-\colorkey{\mathbf{k}_i})\cdot (\mathcal{O}(\colorquery{\mathbf{q}})- \colorvalue{f}(\colorquery{\mathbf{q}}-\colorkey{\mathbf{k}_i};\boldsymbol{\phi}_i)).
    \end{aligned}
\end{equation}
This expression is valid for any choice of $\colorvalue{f}$, not limited to the polynomial function introduced in~\cref{eq:poly-func}. The formulation bears resemblance to~\cref{eq:func-interp}, with $\colorvalue{f}$ replaced by $g$.
The gradient $\frac{\partial \mathcal{O}(\colorquery
{\mathbf{q}})}{\partial \colorquery{\mathbf{q}}}\in\mathbb{R}^3$ is valuable in several contexts. For example, when $\colorquery{\mathbf{q}}$ lies on the surface (zero level-set of $\mathcal{O}(\colorquery{\mathbf{q}})$), $\frac{\partial \mathcal{O}(\colorquery{\mathbf{q}})}{\partial \colorquery{\mathbf{q}}}$ represents the normal direction. Another example is that signed distance functions are expected to satisfy the Eikonal equation, $\left\|\frac{\partial \mathcal{O}(\colorquery{\mathbf{q}})}{\partial \colorquery{\mathbf{q}}}\right\|^2=1$.

\subsection{Extension: Offsets}\label{sec:mean-shift}
Instead of using fixed grid points as \mykeys $\{\mathbf{k}_i\}_i^I$, the points can be allowed to move freely within the space. This approach provides greater flexibility in the representation and can be expressed as:
\begin{equation}\label{eq:func-with-offset}
    \mathcal{O}^\Delta(\colorquery{\mathbf{q}}; \left\{\colorkey{\mathbf{k}_i+\Delta_i}, \beta_i', \boldsymbol{\phi}_i'\right\}_{i=1}^I),
\end{equation}
where the parameter space becomes $\mathbb{R}^{R^3 \times (4+|\boldsymbol{\phi}'|)}$. Importantly, we are still using grids as the data structures.

Another choice is that we can have this combination,
\begin{equation}\label{eq:func-offset}
    \begin{aligned}
        &\mathcal{O}^{+\Delta}(\colorquery{\mathbf{q}}; \left\{\colorkey{\mathbf{k}_i}, \Delta_i, \beta_i, \beta_i', \boldsymbol{\phi}_i, \boldsymbol{\phi}_i'\right\}_{i=1}^I) \\
        =& \mathcal{O}(\colorquery{\mathbf{q}}; \left\{\colorkey{\mathbf{k}_i}, \beta_i, \boldsymbol{\phi}_i\right\}_{i=1}^I\cup \left\{\colorkey{\mathbf{k}_i+\Delta_i}, \beta_i', \boldsymbol{\phi}_i'\right\}_{i=1}^I)
    \end{aligned}
\end{equation}
We aim for $\mathcal{O}$ to represent the bounding space and $\mathcal{O}^{\Delta}$ to capture the near-surface region. In this case, $\{\colorkey{\mathbf{k}_i+\Delta_i}\}_i^I$ should be around the surface. As a result, $\mathcal{O}^{+\Delta}$ effectively reconstructs both the empty regions and the near-surface areas. The parameter space is represented by $\boldsymbol{\theta}\in\mathbb{R}^{R^3 \times (5+|\boldsymbol{\phi}|+|\boldsymbol{\phi}'|)}$. 

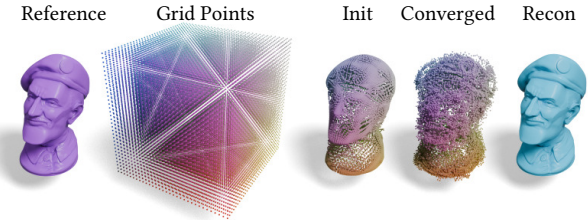
\begin{figure}[tb]
    \begin{overpic}[trim=3cm 0cm 3cm 0cm,clip,width=1\linewidth,grid=false]{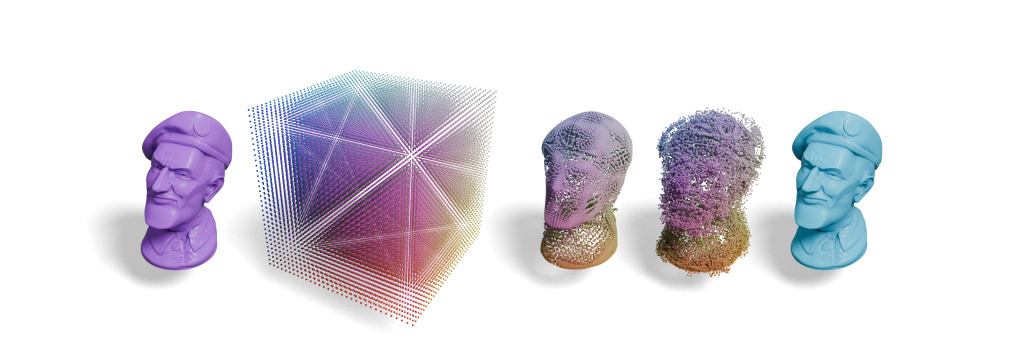}
    \put (7,35) {\small Reference}
    \put (28,35) {\small Grid Points}
    \put (57,35) {\small Init}
    \put (66,35) {\small Converged}
    \put (85,35) {\small Recon}
    \end{overpic}
    \vspace{-30pt}
    \caption{We learn an offset $\Delta_i$ for each grid point $\colorkey{\mathbf{k}_i}$ (\emph{second column}). The initialization (\emph{third column}) is obtained using mean shift. They are around the surface region. After convergence, the offsets moved a little bit (\emph{fourth column}). The final reconstruction is shown on the right (\emph{last column}).}
    \label{fig:pc}
\end{figure}

\paragraph{Initializing Offsets.} 
We observed that during training, the offsets $\Delta_i$ naturally tend to point toward the surface. To accelerate convergence, we propose an initialization strategy inspired by the mean-shift algorithm, which is traditionally used for locating the maxima of a density function. In our context, the surface points represent regions of high density. Therefore, we leverage mean-shift to initialize the offsets $\Delta_i$,
\begin{equation}
    \Delta_i = \frac{\sum^N_n\left[
        \exp\left(-100 \left\|\colorkey{\mathbf{k}_i}-\mathbf{s}_n\right\|^2\right)\cdot\mathbf{s}_n
        \right]}{\sum^N_n \exp\left(-100 \left\|\colorkey{\mathbf{k}_i}-\mathbf{s}_n\right\|^2\right)} - \colorkey{\mathbf{k}_i},
\end{equation}
where $\mathbf{s}_n$ is a surface point and we use $N=16384$ surface points in almost all the experiments. The initialization serves as guidance to identify regions where details are required. Some other works~\cite{ohtake2023mpu,liu2021deep, wang2022dual,he2024dmp} employ octrees to locate the near-surface regions.
An illustration can be found in~\cref{fig:pc}.

\subsection{Fitting}\label{sec:fitting}
We aim to fit SDFs using the proposed representations. The training set is a collection of point-distance pairs, $\{(\colorquery{\mathbf{q}_j}, o_j)\}_j^J$.
The mean squared error (MSE) loss function is minimized as follows:
\begin{equation}\label{eq:loss}
    \mathcal{L}\left(
        \{(\colorquery{\mathbf{q}_j}, o_j)\}_j^J;\boldsymbol\theta
    \right) = \frac{1}{J}\sum_j^J(\mathcal{O}(\colorquery{\mathbf{q}_j}) - o_j)^2.
\end{equation}
In practice, similar to neural networks, we optimize the loss within a small batch size $J_B$ rather than the full dataset size $J$, using stochastic gradient descent methods such as Adam. To avoid notation clutter, we retain the symbol $J$ in subsequent sections.

\subsection{Forward algorithm}\label{sec:forward}
The computation of all the terms $\exp(-\beta_i\left\|\colorquery{\mathbf{q}_j}-\colorkey{\mathbf{k}_i}\right\|^2) f(\colorquery{\mathbf{q}_j}, \colorkey{\mathbf{k}_i};\boldsymbol{\phi}_i)$ ($i=[1,\dots, I]$ and $j=[1,\dots, J]$) requires memory of complexity $\Theta(I\cdot J)$. This poses significant challenges when using PyTorch’s built-in automatic differentiation, particularly with large batch sizes and high-resolution grids. To address this, we developed optimized algorithms for both the forward and backward passes.
A key observation is that the individual terms
$\exp(-\beta_i\left\|\colorquery{\mathbf{q}_j}-\colorkey{\mathbf{k}_i}\right\|^2) f(\colorquery{\mathbf{q}_j}, \colorkey{\mathbf{k}_i};\boldsymbol{\phi}_i)$
do not need to be explicitly stored. Instead, the computation requires only their summation,
$\sum^I_i\exp(-\beta_i\left\|\colorquery{\mathbf{q}_j}-\colorkey{\mathbf{k}_i}\right\|^2) f(\colorquery{\mathbf{q}_j}, \colorkey{\mathbf{k}_i};\boldsymbol{\phi}_i)$. 
By focusing solely on this summation, the memory complexity is reduced to $\Theta(J)$, making the computation more efficient.

The algorithm is summarized in \cref{alg:forward}. Leveraging this approach, we implemented a fused CUDA forward kernel. The summations in the algorithm are further optimized using a fast maximum-reduce operator, supported by modern CUDA architectures. Given the similarity of $\mathcal{O}$ in design to the Attention~\cite{vaswani2017attention} mechanisms, our fused algorithm bears a resemblance to FlashAttention~\cite{dao2022flashattention}, which proposed a fused CUDA kernel for Attention.

\begin{figure}
    \centering
    \includegraphics[width=1\linewidth]{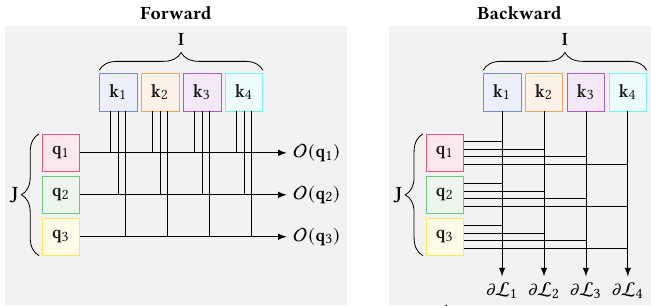}
    \vspace{-20pt}
    \caption{Diagram for the forward and backward pass algorithm.}
    \label{fig:cuda}
\end{figure}

Notably, the gradient $\frac{\partial \mathcal{O}}{\partial \colorquery{\mathbf{q}}}$ in~\cref{eq:func-normal} exhibits a similar structure. As a result, the same algorithm in \cref{alg:forward} is adapted with minimal modifications to compute the gradient efficiently. We show the diagram of the algorithm in~\cref{fig:cuda}.

\begin{algorithm}[tb]
\caption{Forward}\label{alg:forward}
\KwData{$\{\colorkey{\mathbf{k}_i}, \beta_i, \boldsymbol{\phi}_i\}^I_i$ and $\colorquery{\mathbf{q}_j}$}
\KwResult{$\mathcal{O}(\colorquery{\mathbf{q}_j})$, $e_j$}
$e_j=0$\;
$m_j=0$\;
\For{$i\in[1,\cdots, I]$}{
    $l\leftarrow \exp(-\beta_i\left\|\colorquery{\mathbf{q}_j}-\colorkey{\mathbf{k}_i}\right\|^2)$\;
    $e_j\leftarrow e_j + l$ \Comment*[r]{denominator}
    $m_j\leftarrow m_j + l\cdot \colorvalue{f}(\colorquery{\mathbf{q}_j}-\colorkey{\mathbf{k}_i};\boldsymbol{\phi}_i)$ \Comment*[r]{numerator}
}
$\mathcal{O}(\colorquery{\mathbf{q}_j})=m_j/e_j$ \Comment*[r]{save $e_j$ for backward}
\end{algorithm}

\subsection{Backward algorithm}\label{sec:backward}
As discussed earlier, existing automatic differentiation frameworks require substantial memory for this task. To overcome this limitation, the gradients of the loss function
$\mathcal{L}$ with respect to all the parameters $\{\colorkey{\mathbf{k}_i}, \beta_i, \boldsymbol{\phi}_i\}^I_i$ are computed manually. 
By the chain rule, for any loss function $\mathcal{L}$, we have
\begin{equation}
    \begin{aligned}
        \frac{\partial \mathcal{L}}{\partial \colorkey{\mathbf{k}_i}} & = \sum^J_{j=1}\frac{\partial \mathcal{L}}{\partial \mathcal{O}(\colorquery{\mathbf{q}_j})} \frac{\partial \mathcal{O}(\colorquery{\mathbf{q}_j})}{\partial \colorkey{\mathbf{k}_i}}, \\
        \frac{\partial \mathcal{L}}{\partial \beta_i} & = \sum^J_{j=1}\frac{\partial \mathcal{L}}{\partial \mathcal{O}(\colorquery{\mathbf{q}_j})} \frac{\partial \mathcal{O}(\colorquery{\mathbf{q}_j})}{\partial \beta_i}, \\
        \frac{\partial \mathcal{L}}{\partial \boldsymbol\phi_i} & = \sum^J_{j=1}\frac{\partial \mathcal{L}}{\partial \mathcal{O}(\colorquery{\mathbf{q}_j})} \frac{\partial \mathcal{O}(\colorquery{\mathbf{q}_j})}{\partial \boldsymbol\phi_i}.
    \end{aligned}
\end{equation}

\begin{algorithm}[tb]
\caption{Backward}\label{alg:backward}
\KwData{$\{\colorkey{\mathbf{k}_i}, \beta_i, \boldsymbol{\phi}_i\}^I_i$, and $\displaystyle\left\{\colorquery{\mathbf{q}_j}, \mathcal{O}(\colorquery{\mathbf{q}_j)}, e_j, \frac{\partial\mathcal{L}}{\partial\mathcal{O}(\colorquery{\mathbf{q}_j})}\right\}_j^J$}
\KwResult{$\displaystyle\left\{
\frac{\partial \mathcal{L}}{\partial \colorkey{\mathbf{k}_i}},
\frac{\partial \mathcal{L}}{\partial \beta_i},
\frac{\partial \mathcal{L}}{\partial \boldsymbol{\phi}_i}
\right\}_i^I$}
$\partial \mathcal{L}/\partial \colorkey{\mathbf{k}_i}\leftarrow 0$\;
$\partial \mathcal{L}/\partial \beta_i\leftarrow 0$\;
$\partial \mathcal{L}/\partial \boldsymbol{\phi}_i\leftarrow 0$\;
\For{$j\in[1,\cdots, J]$}{
    $l\leftarrow \exp(-\beta_i\left\|\colorquery{\mathbf{q}_j}-\colorkey{\mathbf{k}_i}\right\|^2)/e_j$ \Comment*[r]{softmax}
    $\displaystyle \colorvalue{f}_i \leftarrow \colorvalue{f}(\colorquery{\mathbf{q}}-\colorkey{\mathbf{k}_i};\boldsymbol{\phi}_i)$ \Comment*[r]{evaluate $\colorvalue{f}$}
    $\displaystyle \partial \colorvalue{f}/\partial \colorkey{\mathbf{k}_i} \leftarrow \frac{\partial \colorvalue{f}(\colorquery{\mathbf{q}}-\colorkey{\mathbf{k}_i};\boldsymbol{\phi}_i)}{\partial \colorkey{\mathbf{k}_i}}$ \;
    $\displaystyle \partial \colorvalue{f}/\partial \boldsymbol{\phi}_i \leftarrow \frac{\partial \colorvalue{f}(\colorquery{\mathbf{q}}-\colorkey{\mathbf{k}_i};\boldsymbol{\phi}_i)}{\partial \boldsymbol{\phi}_i}$ \;
    \Comment*[h]{grad w.r.t. $\colorkey{\mathbf{k}_i}$} \;
    $\partial\mathcal{O}/\partial\colorkey{\mathbf{k}_i}\leftarrow l \cdot \left\{\partial \colorvalue{f}/\partial \colorkey{\mathbf{k}_i} + 2\beta_i\left[\colorvalue{f_i}-\mathcal{O}(\colorquery{\mathbf{q}_j})\right](\colorquery{\mathbf{q}_j}-\colorkey{\mathbf{k}_i})\right\}$\;
    $\partial \mathcal{L}/\partial \colorkey{\mathbf{k}_i}\leftarrow \partial \mathcal{L}/\partial \colorkey{\mathbf{k}_i} + \partial\mathcal{L}/\partial\mathcal{O}(\colorquery{\mathbf{q}_j}) \cdot \partial\mathcal{O}/\partial\colorkey{\mathbf{k}_i}$\;
    \Comment*[h]{grad w.r.t. $\beta_i$} \;
    $\partial\mathcal{O}/\partial\beta_i\leftarrow l \cdot \left\{-\left\|\colorquery{\mathbf{q}_j}-\colorkey{\mathbf{k}_i}\right\|^2\left[\colorvalue{f_i}-\mathcal{O}(\colorquery{\mathbf{q}_j})\right]\right\}$\;
    $\partial \mathcal{L}/\partial \beta_i\leftarrow \partial \mathcal{L}/\partial \beta_i + \partial\mathcal{L}/\partial\mathcal{O}(\colorquery{\mathbf{q}_j}) \cdot \partial\mathcal{O}/\partial\beta_i$\;
    \Comment*[h]{grad w.r.t. $\boldsymbol{\phi}_i$} \;
    $\partial\mathcal{O}/\partial\boldsymbol{\phi}_i\leftarrow l \cdot \left\{\partial \colorvalue{f}/\partial \boldsymbol{\phi}_i\right\}$\;
    $\partial \mathcal{L}/\partial \boldsymbol{\phi}_i\leftarrow \partial \mathcal{L}/\partial \boldsymbol{\phi}_i + \partial\mathcal{L}/\partial\mathcal{O}(\colorquery{\mathbf{q}_j}) \cdot \partial\mathcal{O}/\partial\boldsymbol{\phi}_i$\;
}
\end{algorithm}
The gradient of the function $\mathcal{O}$ with respect to the parameters are
\begin{equation}
    \begin{aligned}
        \frac{\partial \mathcal{O}(\colorquery{\mathbf{q}})}{\partial \colorkey{\mathbf{k}_i}} &=&& 
        \colorweight{\mathrm{softmax}}(-\beta_i\left\|\colorquery{\mathbf{q} - \colorkey{\mathbf{k}_i}}\right\|^2)
        \\
        &&& \cdot \left\{\frac{\partial \colorvalue{f}(\colorquery{\mathbf{q}}-\colorkey{\mathbf{k}_i};\boldsymbol{\phi}_i)}{\partial \colorkey{\mathbf{k}_i}}+2\beta_i(\colorquery{\mathbf{q}}-\colorkey{\mathbf{k}_i})\left[\colorvalue{f}(\colorquery{\mathbf{q}}-\colorkey{\mathbf{k}_i};\boldsymbol{\phi}_i))-\mathcal{O}(\colorquery{\mathbf{q}})\right]\right\},
    \end{aligned}
\end{equation}
\begin{equation}
    \begin{aligned}
        \frac{
            \partial \mathcal{O}(\colorquery{\mathbf{q}})
        }{
            \partial \beta_i
        } &=&& 
        \colorweight{\mathrm{softmax}}(-\beta_i\left\|\colorquery{\mathbf{q} - \colorkey{\mathbf{k}_i}}\right\|^2)
        \\
        &&& \cdot \left\{-\left\|\colorquery{\mathbf{q}}-\colorkey{\mathbf{k}_i}\right\|^2 \cdot [\colorvalue{f}(\colorquery{\mathbf{q}}-\colorkey{\mathbf{k}_i};\boldsymbol{\phi}_i)-\mathcal{O}(\colorquery{\mathbf{q}})]\right\},
    \end{aligned}
\end{equation}

\begin{equation}
    \begin{aligned}
        \frac{\partial \mathcal{O}(\colorquery{\mathbf{q}})}{\partial \boldsymbol{\phi}_i} &= &&
        \colorweight{\mathrm{softmax}}(-\beta_i\left\|\colorquery{\mathbf{q} - \colorkey{\mathbf{k}_i}}\right\|^2) 
        \frac{\partial \colorvalue{f}(\colorquery{\mathbf{q}}-\colorkey{\mathbf{k}_i};\boldsymbol{\phi}_i)}{\partial \boldsymbol{\phi}_i}.
    \end{aligned}    
\end{equation}
It is important to note that all calculations require the denominator
$\sum^I_i \exp\left(-\beta_i\left\|\colorquery{\mathbf{q}} -\colorkey{\mathbf{k}_i}\right\|^2\right)$. 
Therefore, this value is directly utilized from the forward pass algorithm. The complete implementation is detailed in~\cref{alg:backward}. Furthermore, all computations are efficiently fused into a single CUDA kernel for optimal performance.


\begin{table*}[]
    \centering
    \caption{Chamfer $\downarrow$ ($\times 10^3$) metrics. The numbers shown in the parenthesis indicate the file size in \texttt{.obj} format. 
    The Chamfer distance is evaluated on 100k surface points. We also show trilinear interpolation ($\mathcal{O}^{\text{trilinear}})$ results as a reference. We highlighted the \colorbox{tablered}{best},\colorbox{orange}{second best}, and \colorbox{lightyellow}{third best}.}
    \label{tab:main}
        \begin{tabular}{cccccccccc}
        \toprule
         & \begin{tabular}{c}
              Type-I\\
              $\mathcal{F}(\mathbf{q})$
         \end{tabular} & 
         \begin{tabular}{c}
              Type-II\\
              $\mathrm{MLP}(\mathbf{q})$
         \end{tabular}
          & \multicolumn{3}{c}{
          \begin{tabular}{c}
              Type-III\\
              $\mathrm{MLP}(\mathcal{F}(\mathbf{q}))$
         \end{tabular}
} & \multicolumn{4}{c}{Ours} \\
         \cmidrule(lr){2-2}\cmidrule(lr){3-3}\cmidrule(lr){4-6}\cmidrule(lr){7-10}
         & Trilinear ($256^3$) & FFN & NGLOD & I-NGP & DiF-Grid & $4^3\times 13$ & $8^3\times 13$ & $16^3\times 13$ & $32^3\times 13$ \\
         \midrule
       Data Structures & Grid & - & Octree & Hash & Grid & Grid & Grid & Grid & Grid \\
      \# Parameters & 16.8m & 0.46m & 10.1m & 14.2m & 5.3m & 832 & 6656 & 53k & 0.43m \\ 

       Non-Neural & \ding{51} & \ding{55} & \ding{55} & \ding{55} & \ding{55} & \ding{51} & \ding{51} & \ding{51} & \ding{51}\\
       \midrule

       \rowcolor{tabcol!10}Armadillo (3.71MB) & 11.964 & 7.625 & 7.420 & \cellcolor{lightyellow}{7.354} & \cellcolor{orange}{7.352} & 14.253 & 8.850& 7.622 &  \cellcolor{tablered}{7.336} \\
       Happy (3.76MB) & 10.531 & 6.942 & 6.858 & \cellcolor{lightyellow}{6.817} & \cellcolor{orange}{6.808} & 19.640 & 9.349 & 7.175 & \cellcolor{tablered}{6.807} \\
       \rowcolor{tabcol!10}Dragon (9.65MB) & 9.514 & 5.856 & 5.722 & \cellcolor{orange}{5.635} & \cellcolor{tablered}{5.610} & 29.952 & 10.405 & 6.228 &  \cellcolor{lightyellow}{5.656}\\
       Lucy (3.71MB) & 9.564 & 5.987 & 5.835 & \cellcolor{lightyellow}{5.734} & \cellcolor{tablered}{5.702} & 18.144 & 9.603 & 6.336 & \cellcolor{orange}{5.717} \\
       \rowcolor{tabcol!10}Fandisk (0.45MB) & 12.847 & 8.581 & 8.493 & \cellcolor{lightyellow}{8.470} & \cellcolor{orange}{8.461} & 14.398 & 8.990 & 8.505 & \cellcolor{tablered}{8.410} \\
       Bed (0.06MB) & 12.281 & 7.315 & 7.364 & \cellcolor{orange}{7.227} & \cellcolor{tablered}{7.213} & 31.442 & 11.084 & 8.190 &  \cellcolor{lightyellow}{7.232} \\
       \rowcolor{tabcol!10}Building (13.67MB) & 17.456 & 12.715 & \cellcolor{lightyellow}{12.710} & \cellcolor{tablered}{12.676} & 12.717 & 31.286 & 15.497 & 12.804 & \cellcolor{orange}{12.685}\\
       Container (1.72MB) & 20.118 &  \cellcolor{lightyellow}{14.069} & \cellcolor{tablered}{13.712} & 14.411 &15.151 & 22.704 & 16.194 & 14.292 & \cellcolor{orange}{13.976}\\
       \rowcolor{tabcol!10}Snowflake (0.22MB) & 19.572 & 15.420 & 15.343 & \cellcolor{orange}{15.268} & \cellcolor{lightyellow}{15.283} &38.916 & 23.407 & 15.776 & \cellcolor{tablered}{15.223}\\
       \midrule
       Mean & 13.761 &	9.390&	\cellcolor{orange}{9.273}	& \cellcolor{lightyellow}{9.288}&	9.366	&24.526	&12.598	&9.659	& \cellcolor{tablered}{9.227} \\
       \bottomrule
    \end{tabular}

\end{table*}
\section{Experiments}
\subsection{Metrics}

\begin{figure*}
    \centering
    \begin{overpic}[trim=1.5cm 3cm 1.5cm 0cm,clip,width=1\linewidth,grid=false]{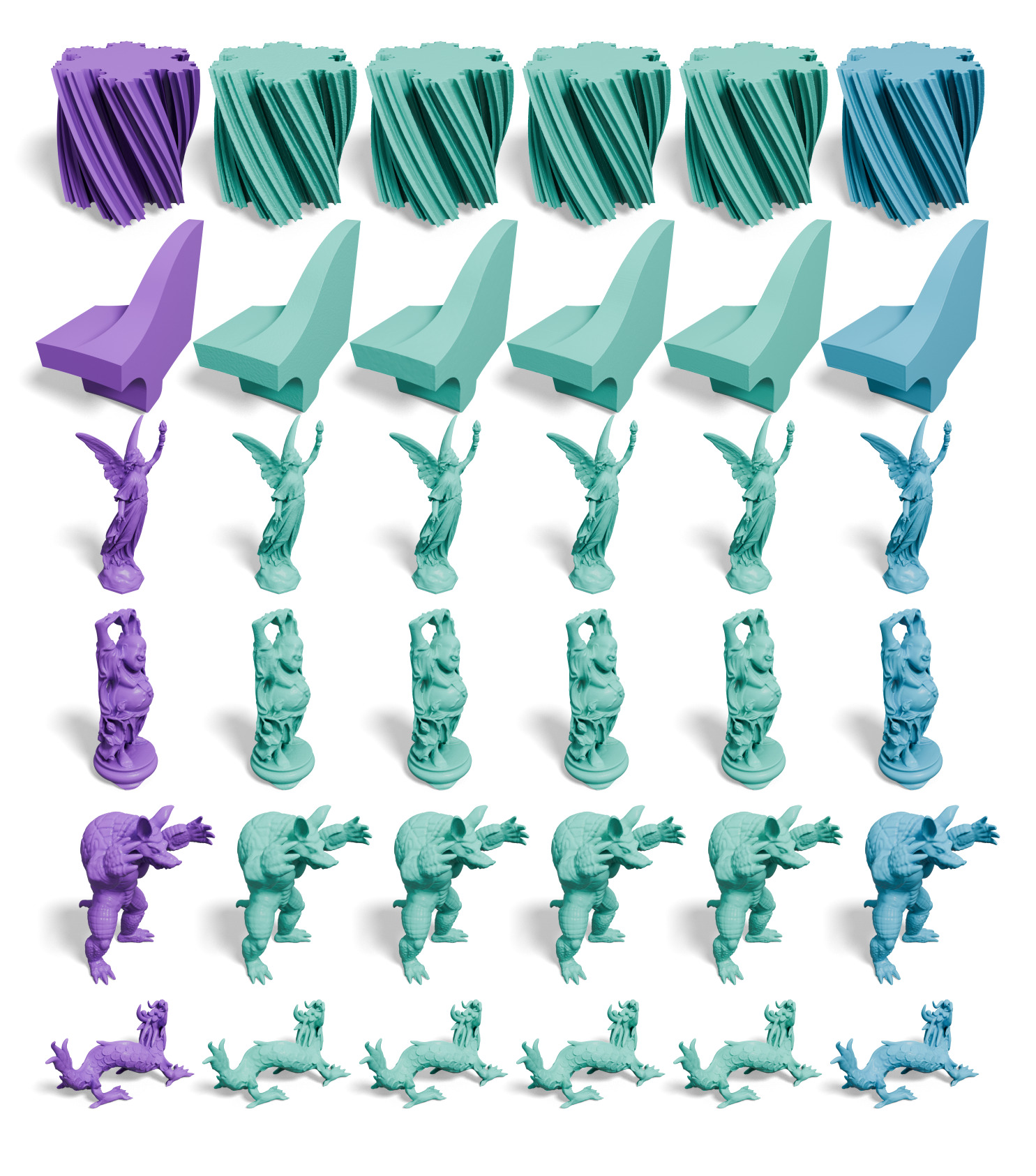}
        \put (6,97) {Reference}    
        \put (21,97) {FFN}    
        \put (34,97) {NGLOD}    
        \put (47,97) {InstantNGP}    
        \put (62,97) {DiF-Grid}    
        \put (74,97) {Ours ($32^3\times 13$)}    
    \end{overpic}
    \caption{We compare the input (left) to the main competitors, FFN~\cite{tancik2020fourier}, NGLOD~\cite{takikawa2021neural}, Instant NGP~\cite{muller2022instant}, Dif-Grid~\cite{chen2023dictionary}, and our representation.}
    \label{fig:main-sga}
\end{figure*}

For easier evaluation we convert SDFs into meshes. To obtain a mesh, we first evaluate $\mathcal{O}(\colorquery{\mathbf{q}})$ at 512-resolution grid points. Then, we use Marching Cubes~\cite{lorensen1998marching} on the resulting grid. We evaluate the Chamfer distance between the reconstructed and ground-truth meshes by sampling 100k points. The Chamfer distance (\texttt{CD}) is defined as,

\begin{equation}
    \begin{aligned}
    \mathrm{CD}(\mathcal{X}_{\mathbf{a}}, \mathcal{X}_{\mathbf{b}}) = & \frac{1}{|\mathcal{X}_{\mathbf{a}}|}
    \sum_{\mathbf{a}\in\mathcal{X}_{\mathbf{a}}}\min_{\mathbf{b}\in\mathcal{X}_{\mathbf{b}}} \left\|\mathbf{a}-\mathbf{b}\right\|_2 + \frac{1}{|\mathcal{X}_{\mathbf{b}}|}
    \sum_{\mathbf{b}\in\mathcal{X}_{\mathbf{b}}}\min_{\mathbf{a}\in\mathcal{X}_{\mathbf{a}}} \left\|\mathbf{a}-\mathbf{b}\right\|_2.
    \end{aligned}
\end{equation}
where $\mathcal{X}_a$ and $\mathcal{X}_b$ means set of samples on shape $a$ and shape $b$, respectively.

While Chamfer distance measures the surface reconstruction quality, we also present metrics related to the function-fitting accuracies, including mean absolute errors (\texttt{AE}) for continuous values and intersection over union (\texttt{IOU}) for binarized values. Both \texttt{AE} and \texttt{IOU} are evaluated on 5 million points sampled in the bounding volume (\texttt{Volume}) and another 5 million points sampled in the near-surface region (\texttt{Near}), resulting in four additional metrics, \texttt{Volume-AE}, \texttt{Volume-IOU}, \texttt{Near-AE} and \texttt{Near-IOU}.



\subsection{Results}
We train the model with AdamW~\cite{loshchilov2017decoupled} optimizer with a learning rate of 6e-4. 
In each iteration, we sample 16384 points in the bounding volume and another 16384 points in the near-surface region.
Since the fused cuda kernel used very little GPU memory, as shown in~\cref{tab:memory_time}, it is also possible to use even larger batch sizes. We experiment on a selected set of meshes (most of them are taken from Thingi10k~\cite{zhou2016thingi10k}). The meshes may include flat surfaces, sharp corners/edges, thin structures, and high-frequency details.

\begin{figure}[tb]
    \centering
    \begin{overpic}[trim=0.5cm 0.5cm 0.5cm 0cm,clip,width=1\linewidth,grid=false]{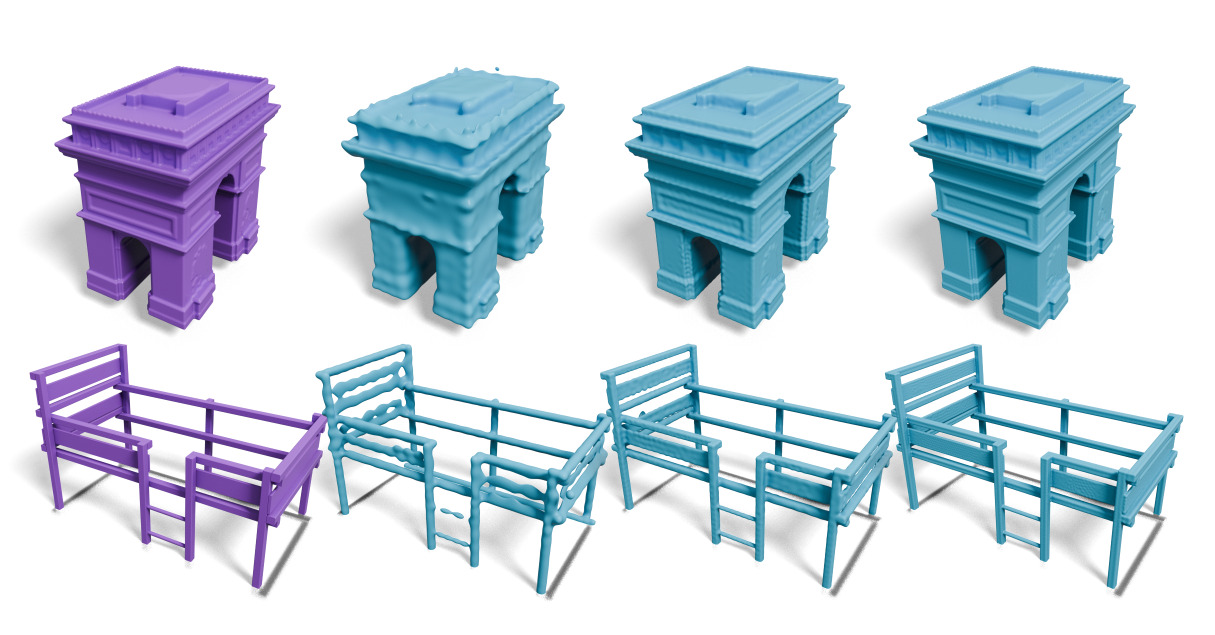}
        \put (5.5,48) {\small{Reference}}
        \put (26.5,48) {\small{Ours ($8^3\times 13$)}} 
        \put (50.5,48) {\small{Ours ($16^3\times 13$)}}    
        \put (75.5,48) {\small{Ours ($32^3\times 13$)}}   
    \end{overpic}
    \vspace{-25pt}
    \caption{\textbf{Ablation study of resolutions.} Results from different variants of our methods are shown in blue. Obviously, when increasing the resolutions, our results exhibit significant quality improvements. The numbers shown in the parenthesis indicate the number of parameters.}
    \label{fig:abl-resolution}
\end{figure}

\paragraph{Main results.} 
The main results are visualized in~\cref{fig:main-sga}. We compare our method against the main competitors, FFN~\cite{tancik2020fourier}, NGLOD~\cite{takikawa2021neural}, Instant NGP~\cite{muller2022instant} and DiF-Grid~\cite{chen2023dictionary} on nine meshes. While several closely related works exist (\eg, DMP~\cite{he2024dmp}, M-SDF~\cite{yariv2024mosaic}), they have been omitted due to the unavailability of publicly accessible code for fair comparison.
Our results are getting better when the resolution is increased. When working with a resolution of 32, the visual quality is comparable to the competitors' results, but our number of parameters is far less.
Our results are comparable or even better according to~\cref{tab:main} in terms of Chamfer distance. Visualizations of other resolutions can be found~\cref{fig:abl-resolution}.

\begin{figure}[tb]
    \begin{overpic}[trim=2cm 0cm 0cm 0cm,clip,width=1\linewidth,grid=false]{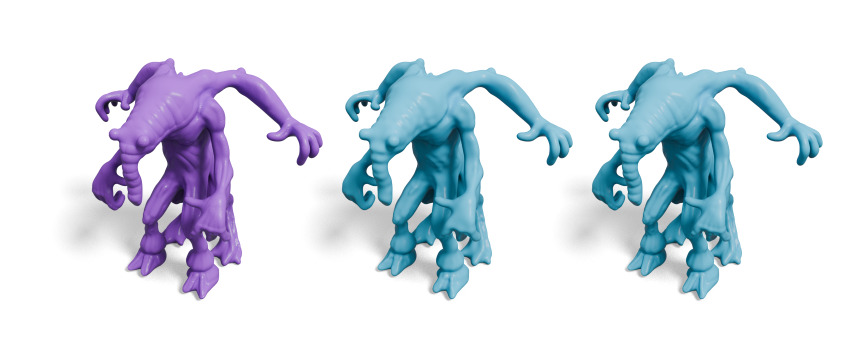}
    \put (10,41) {\small Reference}
    \put (45,41) {\small $\mathcal{O}$}
    \put (74,41) {\small $\mathcal{O}^{+\Delta}$}
    \end{overpic}
    \vspace{-40pt}
    \caption{We show the results of $\mathcal{O}$ (\emph{middle)} and $\mathcal{O}^{+\Delta}$ (\emph{right}). The ground-truth mesh is shown on the \emph{left}. }
    \label{fig:abl-grid-offset}
\end{figure}




\paragraph{Ablation study.} In~\cref{tab:ablation}, we present results analyzing key design choices for the proposed representation. All results were obtained under the same experimental setup.
As we implemented a general framework in addition to our proposed method, our implementation also supports previous methods like~\citet{carr2001reconstruction}, IMLS~\cite{kolluri2008provably} and MPU~\cite{ohtake2023mpu}. We have the following conclusions,
\begin{enumerate}
    \item Learnable \mykeys outperform fixed \mykeys (\eg, \texttt{Config-PC-3} \vs \texttt{Config-PC-1}, \texttt{Config-PC-4} \vs \texttt{Config-PC-2});
    \item Learnable scale parameter ($\beta$) yield better results than fixed scales (\eg, \texttt{Config-G-[5-8]} \vs \texttt{Config-G-[1-4]});
    \item Compared to surface/near-surface \mykeys, grid \mykeys better retain information within the bounding volume, improving \texttt{Volume-AE} and \texttt{Volume-IOU} at the expense of \texttt{Near-AE} and \texttt{Near-IOU} (\eg, \texttt{Config-G-[5-6]} \vs \texttt{Config-PC-[1-2]});
    \item Higher polynomial degrees in \myvalues enhance performance (\eg, from \texttt{Config-G-1} to \texttt{Config-G-4}, from \texttt{Config-G-5} to \texttt{Config-G-8}). However, when the degrees are greater than or equal to 2, visual differences become negligible, as shown in~\cref{fig:intro}.
    \item Larger resolutions improve accuracies (\eg, \texttt{Config-Full-2} \vs \texttt{Config-Full-1}).
    \item The proposed mean-shift initialization (\cref{sec:mean-shift}) is critical for the accuracies (\eg, \texttt{Config-Full-4} \vs \texttt{Config-Full-3}).
    \item Allowing grid \mykeys to be updated during training boosts performance (\eg, \texttt{Config-Full-1} \vs \texttt{Config-G-6}).
    \item While \texttt{Config-Full-1} excels in certain metrics, the \emph{default model} \texttt{Config-Full-4} achieves the best overall performance across surface reconstruction (\texttt{CD}), function values in the bounding volume (\texttt{Volume-AE} and \texttt{Volume-IOU}), and function values in the near-surface region (\texttt{Near-AE} and \texttt{Near-IOU}).
\end{enumerate}


\subsection{Analysis}

\begin{table*}[tb]
    \centering
    \caption{The metrics are CD ($\times 10^3$), AE ($\times 10^4)$, and IOU (\%). \textbf{Grid} means the \mykeys are on a regular grid and not updated during training. \textbf{Off-Grid} means the \mykeys are either surface points or grid points with offsets. \textbf{Deg} denotes the degree of polynomials in~\cref{eq:poly-func}. \textbf{Scale} is the $\beta$ parameter in~\cref{eq:nrbf}.
    We highlighted the \colorbox{tablered}{best},\colorbox{orange}{second best}, and \colorbox{lightyellow}{third best}.}
    \label{tab:ablation}
\def\arraystretch{1}\tabcolsep=0.3em 

\begin{tabular}{l>{\color{textred}}c>{\color{textred}}c>{\color{textpurple}}c>{\color{textpurple}}c>{\color{textblue}}c>{\color{textblue}}c>{\color{textred}}c>{\color{textred}}c>{\color{textpurple}}c>{\color{textpurple}}c>{\color{textblue}}c>{\color{textblue}}cccrccccc}

\toprule
    \multirow{3}{*}{Config} & \multicolumn{6}{c}{Grid} & \multicolumn{6}{c}{Off-Grid} & \multicolumn{3}{c}{Total Params} & \multicolumn{5}{c}{Metrics}\\
    \cmidrule(lr){2-7}\cmidrule(lr){8-13}\cmidrule(lr){14-16}\cmidrule(lr){17-21}
  & \multicolumn{2}{c}{\mykeys} & \multicolumn{2}{c}{\myvalues}       & \multicolumn{2}{c}{\myweights}       & \multicolumn{2}{c}{\mykeys} & \multicolumn{2}{c}{\myvalues}       & \multicolumn{2}{c}{\myweights}  & \multirow{2}{*}{Len} & \multirow{2}{*}{Ch} & \multirow{2}{*}{\# Params} & 
  \multirow{2}{*}{CD $\downarrow$ } & \multicolumn{2}{c}{Volume} & \multicolumn{2}{c}{Near}\\
  \cmidrule(lr){2-3}\cmidrule(lr){4-5}\cmidrule(lr){6-7}\cmidrule(lr){8-9}\cmidrule(lr){10-11}\cmidrule(lr){12-13}
  \cmidrule(lr){18-19}\cmidrule(lr){20-21}
  & Data & \# & Deg & \# & Scale   & \# & Data & \# & Deg & \# & Scale   & \# &  &  &  &              & AE $\downarrow$  & IOU $\uparrow$  & AE $\downarrow$  & IOU $\uparrow$  \\
  \midrule
  \midrule

\rowcolor{tabcol!10}PC-1$^*$ & -    & -      & -      & -      & -       & -      & F    & 3      & 0      & 1      & L       & 1      & 32768  & 5        & 163{,}840 & 13.001          & 85.450         & 99.611          & 16.199         & 89.068          \\
PC-2$^\dag$ & -    & -      & -      & -      & -       & -      & F    & 3      & 1      & 4      & L       & 1      & 32768  & 8        & 262{,}144 & 11.192          & 27.857         & 99.920          & 5.697          & 95.690          \\
\rowcolor{tabcol!10}PC-3 & -    & -      & -      & -      & -       & -      & L    & 3      & 0      & 1      & L       & 1      & 32768  & 5        & 163{,}840 & 11.159          & 23.263         & 99.903          & 6.865          & 94.805          \\
PC-4 & -    & -      & -      & -      & -       & -      & L    & 3      & 1      & 4      & L       & 1      & 32768  & 8        & 262{,}144 & 11.095          & 17.482         & 99.924          & 5.434          & 95.883          \\
\midrule
\rowcolor{tabcol!10}G-1 & F    & 0      & 0      & 1      & F       & 0      & -    & -      & -      & -      & -       & -      & $32^3$  & 1        & 32{,}768  & 22.291          & 119.286        & 98.926          & 50.642         & 62.104          \\
G-2 & F    & 0      & 1      & 4      & F       & 0      & -    & -      & -      & -      & -       & -      & $32^3$  & 4        & 131{,}072 & 18.638          & 38.890         & 99.552          & 21.115         & 83.951          \\
\rowcolor{tabcol!10}G-3 & F    & 0      & 2      & 10     & F       & 0      & -    & -      & -      & -      & -       & -      & $32^3$  & 10       & 327{,}680 & 13.987          & 33.813         & 99.737          & 15.343         & 88.531          \\
G-4 & F    & 0      & 3      & 20     & F       & 0      & -    & -      & -      & -      & -       & -      & $32^3$  & 20       & 655{,}360 & 13.969          & 33.789         & 99.739          & 15.315         & 88.553          \\
\midrule

\rowcolor{tabcol!10}G-5 & F    & 0      & 0      & 1      & L       & 1      & -    & -      & -      & -      & -       & -      & $32^3$  & 2        & 65{,}536  & 12.807          & 29.989         & 99.634          & 22.525         & 82.614          \\
G-6 & F    & 0      & 1      & 4      & L       & 1      & -    & -      & -      & -      & -       & -      & $32^3$  & 5        & 163{,}840 & 11.624          & 16.770         & 99.822          & 12.112         & 90.681          \\
\rowcolor{tabcol!10}G-7$^\ddagger$ & F    & 0      & 2      & 10     & L       & 1      & -    & -      & -      & -      & -       & -      & $32^3$  & 11       & 360{,}448 & 11.514          & 15.798         & 99.840          & 11.027         & 91.546          \\
G-8 & F    & 0      & 3      & 20     & L       & 1      & -    & -      & -      & -      & -       & -      & $32^3$  & 21       & 688{,}128 & 11.524          & 15.797         & 99.839          & 11.020         & 91.553          \\
\midrule
\midrule
\rowcolor{tabcol!10}Full-1$^\circ$ &  -    &    -    &    -    &     -   &     -    &    -    & L    & 3      & 1      & 4      & L       & 1      & $32^3$  & 8        & 262{,}144 & \cellcolor{orange}11.053 & \cellcolor{orange}8.790 & \cellcolor{lightyellow}99.937 & \cellcolor{lightyellow}4.633 & \cellcolor{lightyellow}96.779 \\
Full-2 &  -    &    -    &    -    &     -   &     -    &    -    & L    & 3      & 1      & 4      & L       & 1      & $64^3$  & 8        & 2{,}097{,}152	& \cellcolor{tablered}10.996 &	\cellcolor{tablered}4.453	& \cellcolor{tablered}99.967 &	\cellcolor{tablered}2.118	& \cellcolor{tablered}98.360 \\
\rowcolor{tabcol!10}Full-3 & F    & 0      & 1      & 4      & L       & 1      & L    & 3      & 1      & 4      & L       & 1      & $32^3$  & 13       & 425{,}984 & 11.145          & 12.342         & 99.912          & 6.499          & 95.218          \\
\midrule
Full-4$^\diamond$ & F    & 0      & 1      & 4      & L       & 1      & I+L  & 3      & 1      & 4      & L       & 1      & $32^3$  & 13       & 425{,}984 & \cellcolor{lightyellow}11.057 & \cellcolor{lightyellow}9.079 & \cellcolor{orange}99.947 & \cellcolor{orange}3.772 & \cellcolor{orange}97.119 \\
\bottomrule
\multicolumn{21}{l}{\textbf{F}: Fixed, not updated during training; \textbf{L}: Learnable, updated during training; \textbf{I}: Initialization, mesh-shift initialization.} \\
\multicolumn{21}{l}{$^*$\textbf{Config-PC-1}: (roughly) equivalent to~\citet{carr2001reconstruction}.}\\
\multicolumn{21}{l}{$^\dag$\textbf{Config-PC-2}: (roughly) equivalent to IMLS~\cite{kolluri2008provably}.}\\
\multicolumn{21}{l}{$^\ddagger$\textbf{Config-G-7}: (roughly) equivalent to MPU~\cite{ohtake2023mpu} with only one level.}\\
\multicolumn{21}{l}{$^\circ$\textbf{Config-Full-1}: $\mathcal{O}$ in~\cref{eq:func-interp} with learnable \mykeys; $^\diamond$\textbf{Config-Full-4}: \emph{default} model $\mathcal{O}^{+\Delta}$ in~\cref{eq:func-offset}.}\\
\end{tabular}
\end{table*}

\begin{figure}[tb]
    \begin{overpic}[trim=2cm 0cm 2cm 0cm,clip,width=1\linewidth,grid=false]{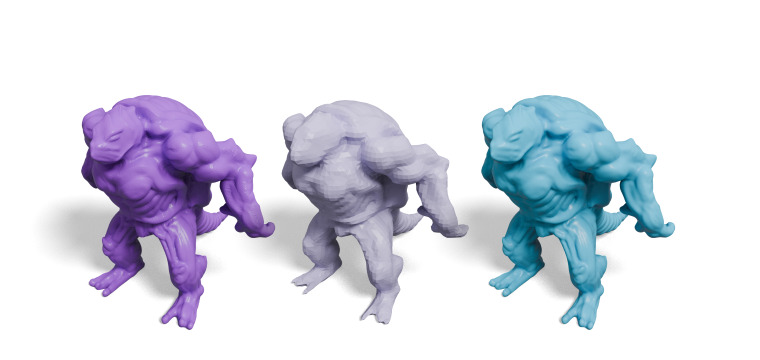}
    \put (10,43) {\small Reference}
    \put (35,44) {\begin{tabular}{c}  
        \small $\mathcal{O}^{\text{trilinear}}$ \\ \small $76^3=438976$
    \end{tabular}}
    \put (63,44) {\begin{tabular}{c}  
        \small $\mathcal{O}^{+\Delta}$ \\ \small $32^3\times 13=425984$
    \end{tabular}}
    \end{overpic}
    \vspace{-35pt}
    \caption{We compare our results against a non-neural representation using trilinear interpolation (\emph{middle}). We choose a resolution of 76 which results in 438976 floats. This is close to our representation (425984). However, our quality (\emph{right}) is significantly better.}\label{fig:comp-trilinear}
\end{figure}

\paragraph{Non-neural representations.} We compare our results with a trilinear grid at a resolution of 76, which uses $76^3=438,976$ floats as the parameters. In contrast, our $\mathcal{O}^{+\Delta}$ representation uses $32^3\times 13=425,984$ floats. Despite utilizing a comparable number of floats, our method significantly outperforms the trilinear grid in quality, as illustrated in~\cref{fig:comp-trilinear}.

\begin{table}[tb]
    \centering
    \caption{We measure the time and memory consumption of our implementation for different resolutions. We compare it against a naive PyTorch implementation. Our method takes much less memory and requires shorter forward and backward times. PyTorch raises an out-of-memory error for 64 resolution.}
    \label{tab:memory_time}
    
\def\arraystretch{1}\tabcolsep=0.3em 
\begin{tabular}{cccccc}

\toprule
$J=16384$ & $I=4^3$  & $I=4^3$ & $I=16^3$ & $I=32^3$ & $I=64^3$ \\ \midrule
py forward (ms)     & 3.259 & 3.843 & 8.341  & 42.209  & -         \\
\rowcolor{tabcol!10}our forward (ms)     & 0.129 & 0.185 & 0.553  & 3.064   & 22.576   \\
py backward (ms)    & 6.143 & 7.086 & 18.910 & 121.037 &  -        \\
\rowcolor{tabcol!10}our backward (ms)     & 0.896 & 1.046 & 2.321  & 10.561  & 82.097   \\ \midrule
py memory     & 0.07G & 0.43G & 3.33G  & 26.64G  & OOM      \\
\rowcolor{tabcol!10}our memory      & 0.62M & 0.63M & 0.70M  & 1.68M   & 10.44M   \\
\bottomrule
\end{tabular}
\end{table}
\paragraph{Running time and memory.} We compare the proposed implementation with PyTorch's~\cite{paszke2019pytorch} built-in automatic differentiation framework. Because of the $\Theta(I\times J)$ complexity, it is challenging to work with large resolutions. According to~\cref{tab:memory_time}, our implemented CUDA kernel uses only 1/10 of memory and 1/10 forward/backward pass time. The efficiency not only lowers resource consumption but also enables training on low-tier desktop GPUs. Querying $256\times 256 \times 256\approx 1.67\times 10^7$ points at the same time takes  only about 5 seconds. Both the training and testing memory usage compared with other methods can be found in~\cref{tab:mem-ingp}. The significantly smaller memory usage suggests that the method can be trained using desktop-level GPUs.

\begin{table}[tb]
    \centering
    \caption{Memory comparison of competitors and ours. The testing memory is evaluated on 16384 query points.}
    \label{tab:mem-ingp}
    \begin{tabular}{cccccc}
    \toprule
        Memory (MB) & FFN & NGLOD & I-NGP & DiF-Grid & Ours\\
         \midrule
        Train  & 327 & 962 & 305 & 243 & 16 \\
        \rowcolor{tabcol!10} Test  & 59 & 124 &54 & 30 &1.68 \\
        \bottomrule
    \end{tabular}
\end{table}

\begin{figure}[tb]
    \begin{overpic}[trim=2cm 0cm 2cm 0cm,clip,width=1\linewidth,grid=false]{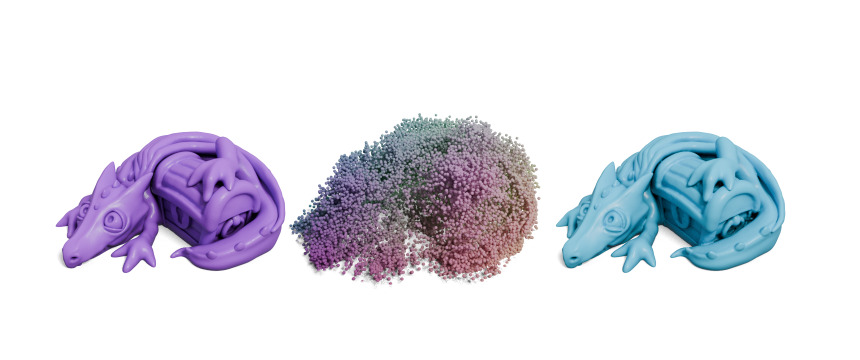}
    \put (9,33) {\small Reference}
    \put (44,33) {\small $\mathbf{k}_i+\Delta_i$}
    \put (78,33) {\small Reconstruction}
    \end{overpic}
    \vspace{-40pt}
    \caption{We visualize the deformed points $\mathbf{k}_i+\Delta_i$ in the middle. The reference and the reconstructed mesh are shown on the left and right, respectively.}
    \label{fig:treasure-pc}
\end{figure}

\paragraph{Deformed grids.}
In the previous section, we argued that the combined representation $\mathcal{O}^{+\Delta}$ gives better performance than $\mathcal{O}$. We visualize the learned deformed grid in~\cref{fig:treasure-pc}.

\begin{figure}
    \centering
    \includegraphics[width=1\linewidth]{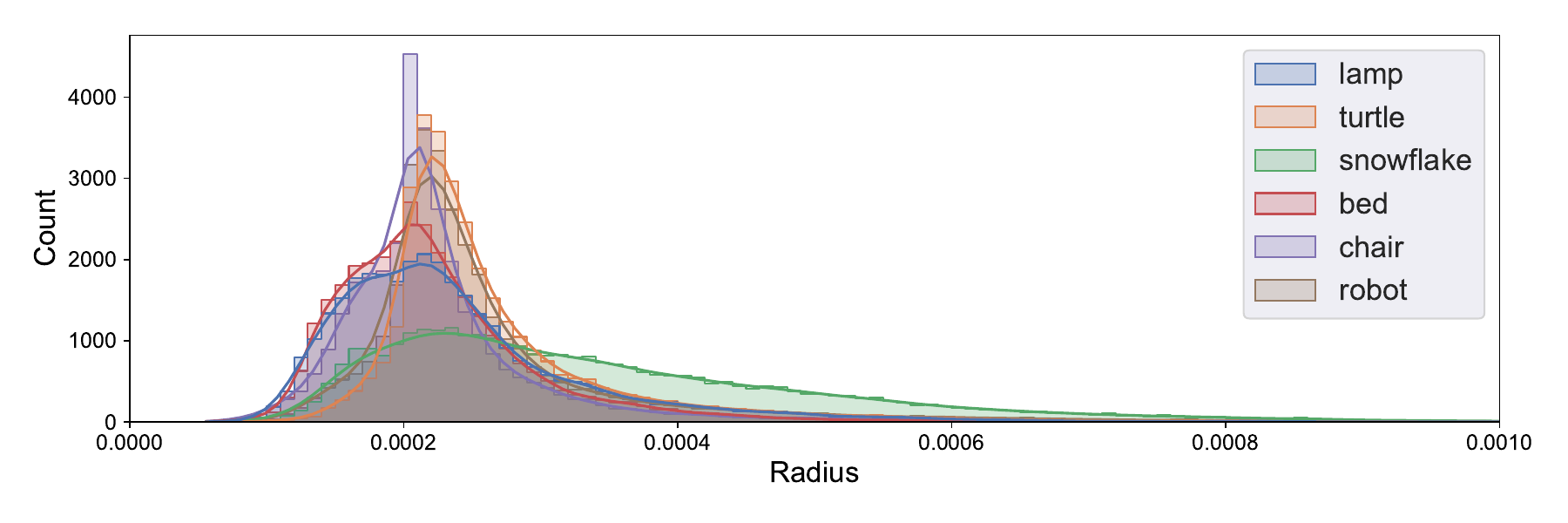}
    \vspace{-20pt}
    \caption{\textbf{Histogram of radius.} Visualizations of the objects can be found in~\cref{fig:teaser}. Different objects have different distributions.}
    \label{fig:radius-count}
\end{figure}

\paragraph{Scales/radius.} 
The inverse of the scale parameter $\beta_i$ represents the radius of each \mykey $\colorkey{\mathbf{k}_i}$. We initialize all scale parameters to $\exp(7)$, which gives a radius of $1 / \exp(7) \approx 0.009$). These parameters are updated during training. After convergence, their distribution may differ significantly from initialization. We visualize the radius (inverse of the scale parameter) in~\cref{fig:radius-count}. We found out that, objects with thin structures (\eg, \texttt{lamp, bed}) exhibit many small radii compared to other objects. The radius parameters are automatically tuned through optimization, which is an advantage of the global support over the manually tuned radius parameter used in local support methods.

\subsection{Application: function decomposition}
\begin{figure}[tb]
    \begin{overpic}[trim=0cm 0cm 0cm 0cm,clip,width=1\linewidth,grid=false]{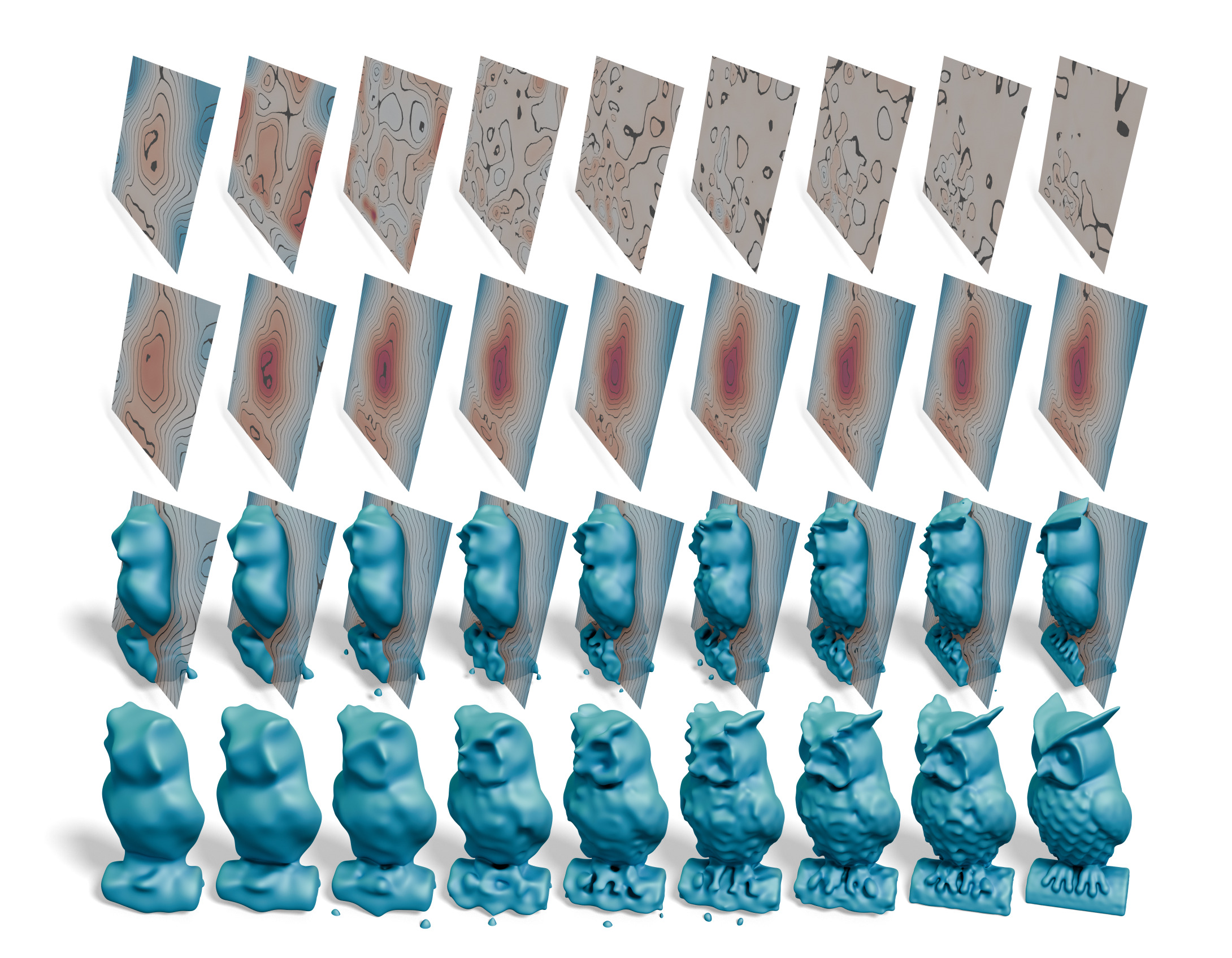}
        \put (10,75) {\scriptsize $b=0$}
        \put (20,75) {\scriptsize $b=1$}
        \put (29,75) {\scriptsize $b=2$}
        \put (38.5,75) {\scriptsize $b=3$}
        \put (48,75) {\scriptsize $b=4$}
        \put (57.5,75) {\scriptsize $b=5$}
        \put (66.5,75) {\scriptsize $b=6$}
        \put (76,75) {\scriptsize $b=7$}
        \put (85.5,75) {\scriptsize $b=8$}
        \put (3, 59) {\small \rotatebox{90}{Each band}}
        \put (3, 39) {\small \rotatebox{90}{Partial sum}}
        \put (3, 10) {\small \rotatebox{90}{Reconstruction}}
        
    \end{overpic}
    \vspace{-25pt}
    \caption{\textbf{Function frequency decomposition.} We show each frequency band (slicing plane) $\colorvalue{\mathcal{O}_{b}}(\colorquery{\mathbf{q}})$ in the \emph{top} row. The partial sum $\sum_{b'=0}^b \colorweight{\cos}(b'\pi\colorquery{\mathbf{q}})\colorvalue{\mathcal{O}_{b'}}(\colorquery{\mathbf{q}})$ of~\cref{eq:cosine-series} (slicing plane) is shown in the \emph{second} row. In \emph{third} and \emph{final} row, we visualize the reconstructed partial sum.}
    \label{fig:freqs}
\end{figure}
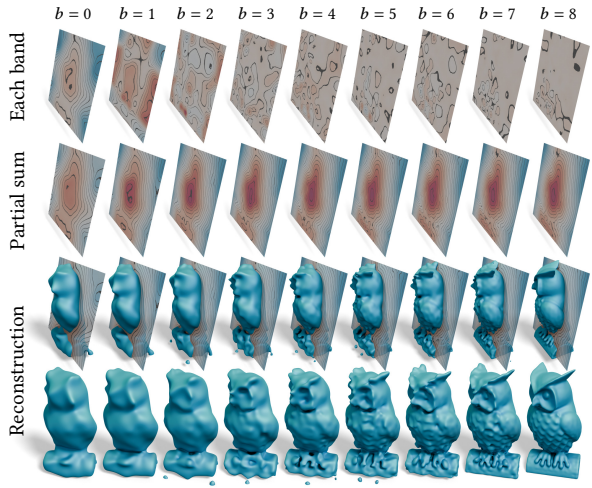

As discussed in~\cref{eq:func-approx-func}, we can use the proposed $\mathcal{O}$ in~\cref{eq:nrbf} as the \myvalue, 
\begin{equation}\label{eq:o-as-value}
    \mathcal{S}(\textcolor{textgreen}{\mathbf{q}}) = \sum_{b} \textcolor{textblue}{w}(\textcolor{textgreen}{\mathbf{q}}, \colorkey{\mathbf{k}_b})
     \cdot \colorvalue{\mathcal{O}_b}(\colorquery{\mathbf{q}}).
\end{equation}
We pick cosine function as the \myweights,
\begin{equation}\label{eq:cosine-series}
    \mathcal{S}(\textcolor{textgreen}{\mathbf{q}}) = 
    \sum_{b=0}^B \colorweight{\cos}(\colorkey{b} \cdot \pi \cdot \colorquery{\mathbf{q}})
     \cdot \colorvalue{\mathcal{O}_b}(\colorquery{\mathbf{q}}),
\end{equation}
which is exactly the form of the Fourier cosine series. When using 16 as the resolution and 1 as the polynomial degree for each $b\in\{0, 1, \dots, B\}$, $\colorvalue{\mathcal{O}_b}$ is parameterized by $16\times 16 \times 16 \times 4$ floats. Thus, the total number of parameters is $20480\times B$. When $B=1$, the setup is equivalent to \texttt{Config-G-6} in~\cref{tab:ablation} with resolution 16. We optimize the function $\mathcal{S}$ as in~\cref{eq:loss}.
The performance is improving when we use more frequency bands.
Visual results can be found~\cref{fig:freqs}. When more frequency bands are used in the reconstruction, the reconstruction surfaces get better.



\subsection{Application: estimating normals}
\begin{figure}[tb]
    \centering
    \includegraphics[width=1\linewidth]{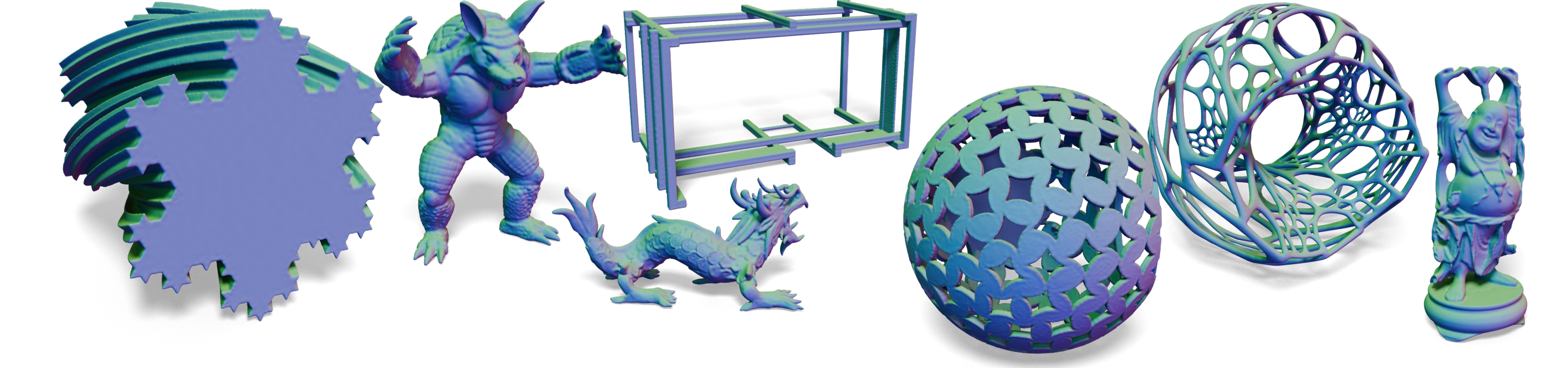}
    \vspace{-25pt}
    \caption{\textbf{Normal maps.} The meshes are obtained using Marching Cubes and then we query the vertices in~\cref{eq:func-normal} to get per-vertex normals.}
    \label{fig:normal-all}
\end{figure}

We can easily calculate the normals using~\cref{eq:func-normal}. 
For ours, we reconstruct the mesh in our representation and obtain the normals using a single forward pass. A visualization is shown in~\cref{fig:normal-all}.

\subsection{Application: shape manipulation}

\begin{figure}[tb]
    \begin{overpic}[trim=0cm 0cm 0cm 0cm,clip,width=1\linewidth,grid=false]{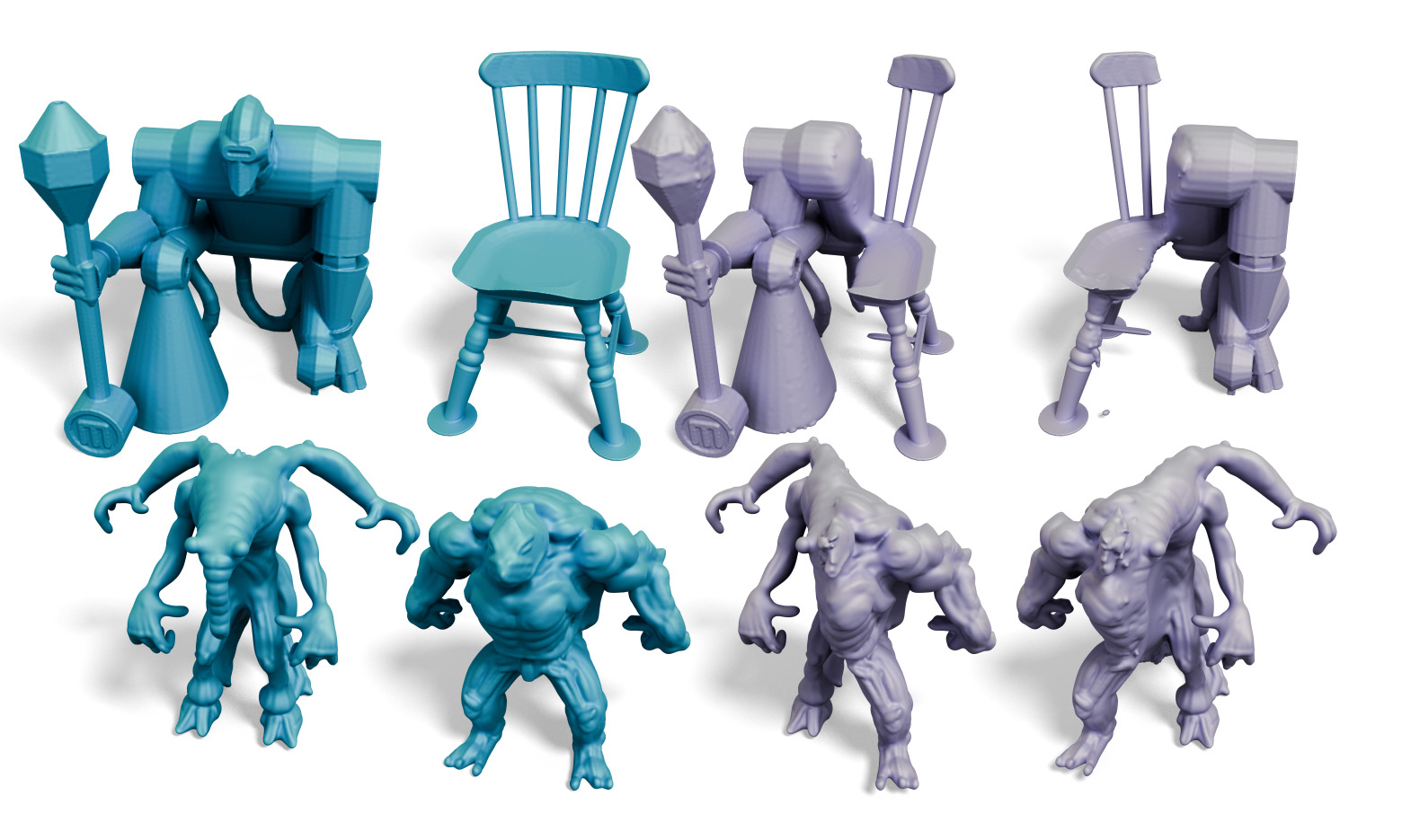}
        \put (12,58) {\small Shape A}
        \put (32,58) {\small Shape B}
        \put (48,58) {\small A left + B right}
        \put (73,58) {\small B left + A right}

    \end{overpic}
    \vspace{-25pt}
    \caption{\textbf{Function manipulation.} Given two shapes (Shape A and Shape B), we can combine them into one by using only half of each shape representation.}
    \label{fig:mixed}
\end{figure}
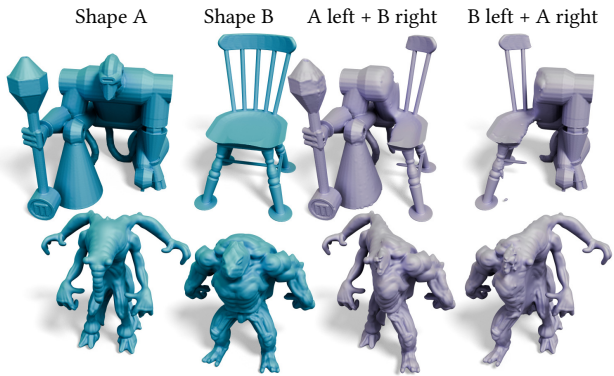

The proposed representations are stored as regular grids, enabling intuitive function manipulation through direct grid operations. As demonstrated in~\cref{fig:mixed}, our framework supports seamless combination of two signed distance functions (SDFs)—a task that highlights its interpretability advantage over purely implicit MLP-based representations (Type-II). Unlike Type-II methods, which encode geometry in abstract network weights and lack explicit spatial control mechanisms, our grid-based approach provides both mathematical transparency and user-editable primitives.

\section{Conclusion}

We introduced a general framework for function representations and a novel approach for encoding 3D signed distance functions (SDFs).
The representation leverages polynomials interpolated by radial basis functions, resulting in a simple yet powerful form. This simplicity enables the use of significantly fewer parameters compared to state-of-the-art representations based on neural networks, octrees, and hashing grids, while achieving comparable approximation quality.

For future work, we aim to extend this representation to 3D geometry generative modeling, including applications with diffusion models and autoregressive generative modeling. Additionally, we plan to explore its applicability to 2D and 4D data.

\paragraph{Limitation.} We would like to discuss two main limitations. 1) Both the training and inference algorithms are slower than those of Instant NGP. As a highly optimized library, Instant NGP can fit a model under 5 minutes, while we still need around 20 minutes.
While we implemented our code in CUDA, we believe further acceleration will be possible by experimenting with various code acceleration strategies. 2) The implementation is largely dependent on GPU architecture (e.g., the maximum number of warps/threads/blocks). Due to these hardware constraints, the current implementation cannot scale to $128^3$ resolution.


\bibliographystyle{ACM-Reference-Format}
\bibliography{sample-bibliography}


\newpage

\end{document}